\documentclass[aps,pre,preprint,groupedaddress,showpacs]{revtex4-1}
\usepackage{amssymb}

\usepackage[dvips]{graphicx}
\usepackage{textcomp}
\usepackage{amssymb}

\begin{document}

\title{\bf The thermodynamic Casimir force: A Monte Carlo study of the 
 crossover between the ordinary and the normal surface universality
           class}

\author{Martin Hasenbusch}
\email[]{Martin.Hasenbusch@physik.hu-berlin.de}
\affiliation{
Institut f\"ur Physik, Humboldt-Universit\"at zu Berlin,
Newtonstr. 15, 12489 Berlin, Germany}

\date{\today}

\begin{abstract}
We study the crossover from the ordinary to the normal surface universality 
class in the three-dimensional Ising bulk universality class. This crossover
is relevant for the behavior of films of binary mixtures near the demixing 
point and a weak adsorption at one or both surfaces.
We perform Monte Carlo simulations of the improved Blume-Capel model on the 
simple cubic lattice. We consider systems with film geometry, where various 
boundary conditions are applied.
We discuss corrections to scaling that are caused by the 
surfaces and their relation with the so called extrapolation length.
To this end we analyze the behavior of the magnetization profile near the 
surfaces of films. 
We obtain an accurate estimate of the renormalization group exponent 
$y_{h_1}=0.7249(6)$ for the ordinary surface universality class.
Next we study the thermodynamic Casimir force in the crossover region 
from the ordinary to the normal surface universality class.  To this 
end, we compute the Taylor-expansion of the crossover finite size 
scaling function
up to the second order in $h_1$ around $h_1=0$, where $h_1$ is the external 
field at one of the surfaces. We check the range of applicability of the 
Taylor-expansion by simulating at finite values of $h_1$.  Finally we 
study the approach to the strong adsorption limit  $h_1 \rightarrow \infty$. 
Our results confirm the qualitative picture that emerges from exact 
calculations for stripes of the two-dimensional Ising model, 
[D. B. Abraham and A. Macio\l ek, Phys. Rev. Lett. {\bf 105}, 055701 (2010)],
mean-field calculations and preliminary Monte Carlo simulations of
the Ising model on the simple cubic lattice, 
[T. F. Mohry et al, Phys.\ Rev.\ E {\bf 81}, 061117 (2010)]: For certain 
choices of $h_1$ and the thickness of the film, the thermodynamic 
Casimir force changes sign as a function of the temperature and for certain 
choices of the temperature and $h_1$, it also changes sign as a function
of the thickness of the film. 
\end{abstract}
\maketitle

\section{Introduction}
In 1978  Fisher and de Gennes \cite{FiGe78} realized that when thermal
fluctuations are restricted by a container, a force acts on its walls.
Since this effect is  analogous to the Casimir effect \cite{Casimir},
where the restriction of quantum fluctuations induces a force, it is called
``thermodynamic'' Casimir effect. Since thermal fluctuations only extend to
large scales in the neighborhood of continuous phase transitions it is
also called ``critical'' Casimir effect. Recently this force could be
detected for various experimental systems and quantitative predictions could
be obtained from Monte Carlo simulations of spin models \cite{Ga09}.

The behavior of the thermodynamic Casimir force can be described by 
finite size scaling (FSS) \cite{Barber} laws. For the film geometry that we
consider here, one gets \cite{Krech} for the thermodynamic Casimir force 
per area
\begin{equation}
\label{FCffs}
 F_{Casimir} \simeq k_B T L_0^{-3} \; 
\theta_{(UC_1,UC_2)}(t [L_0/\xi_0]^{1/\nu}) 
\end{equation}
where $L_0$ is the thickness of the film and $t=(T-T_c)/T_c$ is the reduced
temperature and $T_c$ the critical temperature. 
Note that below, analyzing our data, we shall use for simplicity 
the definition $t=\beta_c-\beta$, where $\beta=1/k_B T$.
The amplitude $\xi_{0}$ of the correlation length $\xi$ is defined by
\begin{equation}
\label{xipower}
\xi = \xi_{0,\pm} |t|^{-\nu} \times (1 + a_{\pm} |t|^{\nu \omega} + c t + ...)
\end{equation}
where $-$ and $+$ indicate the high and the low temperature phase, respectively.
Since the correlation length can be determined more accurately in the high 
temperature phase than in the low temperature phase, we take $\xi_{0}=\xi_{0,+}$
in eq.~(\ref{FCffs}). The power law~(\ref{xipower}) is subject to
confluent corrections, such as $a_{\pm} |t|^{\nu \omega}$, and non-confluent 
ones such as $c t$. Critical exponents such as $\nu$ and ratios of 
amplitudes such as $\xi_{0,+}/\xi_{0,-}$ are universal. Also correction
exponents such as $\omega$ and ratios of correction
amplitudes such as $a_+/a_-$ are universal. For the three-dimensional Ising
universality class, which is considered here $\nu \omega \approx 0.5$. 
For reviews on critical phenomena and their modern theory, i.e.,
the Renormalization Group (RG) see, e.g., \cite{WiKo,Fisher74,Fisher98,PeVi02}.
The universal finite size scaling
function $\theta_{(UC_1,UC_2)}$ depends on the universality class of the 
bulk system as well as the surface universality classes $UC_1$ and $UC_2$ of 
the two surfaces of the film. For reviews on  surface critical phenomena
see e.g.  \cite{BinderS,Diehl86,Diehl97}. We shall give a brief discussion 
below in section \ref{semiinfite}. 

In the past few years there has been great interest in the crossover behaviors 
of the thermodynamic Casimir force. In \cite{ScDi08} the authors have studied 
the crossover from the special surface universality class to the ordinary 
one by using field theoretic methods. They find that for certain choices of the
parameters, the thermodynamic Casimir force changes sign with a varying 
thickness of 
the film. The authors of \cite{AbMa09} have computed exactly the thermodynamic 
Casimir force for stripes of the two-dimensional Ising model as a function
of the external surface fields $h_1$ and $h_2$. Also here the authors have
found that for certain choices of the fields $h_1$ and $h_2$, the thermodynamic
Casimir force does change sign as a function of the temperature or the 
thickness of the film.
More recently, the authors of \cite{MoMaDi10} have studied 
the crossover from the ordinary to the normal surface universality class, 
and the crossover from the special to the ordinary as well as the 
normal surface universality class using the mean-field approximation.
Also in these cases a change of sign of the thermodynamic Casimir force 
could be observed.
Furthermore in \cite{MoMaDi10} preliminary results \cite{Va} of 
Monte Carlo simulations of the spin-1/2 Ising model on the simple cubic 
lattice for the crossover from the ordinary to the normal surface universality
class were presented.
Following the authors of \cite{MoMaDi10} these observations might be of 
technological relevance. They write: 
"Such a tunability of critical Casimir forces towards repulsion might 
be relevant for micro- and nano-electromechanical systems in order to prevent 
stiction due to the omnipresent attractive quantum mechanical Casimir forces
\cite{Casimir,Ball}."
In recent experiments on colloidal particles immersed in a binary mixture of
fluids \cite{NeHeBe09}, the authors have demonstrated that the adsorption
strength can be varied continuously by a chemical modification of the
surfaces. In particular the situation of effectively equal adsorption
strengths for the two fluids can be reached. For sufficiently small
ordering interaction at the surface, this corresponds to the ordinary
surface universality class.  Hence these experiments open the way to
study the crossover from the ordinary to the normal universality class.
%%P1
As discussed in refs. \cite{PAT1,PAT2,PAT3,PAT4} effectively weak 
adsorption can also be obtained by using patterned substrates. 

In the present work we compute scaling functions for the film or plate-plate
geometry.
In order to compare with experiments on the thermodynamic Casimir force 
between colloidal particles and a flat substrate as studied in ref.
\cite{NeHeBe09} the scaling function for the plate-sphere geometry has to be 
computed.
The Derjaguin approximation \cite{De34} might be used to derive scaling 
functions for the plate-sphere geometry from those for the plate-plate geometry
if the radius of the sphere is large compared with the distance between
the plate and the sphere \cite{HaScEiDi98,ScHaDi03}, as it is indeed the 
case in ref. \cite{NeHeBe09}. In the recent works \cite{Nature08,Gaetal09} 
the Derjaguin approximation had been used to obtain the scaling 
functions for the plate-sphere geometry in the strong adsorption limit 
starting from the Monte Carlo estimates of refs. \cite{VaGaMaDi07,VaGaMaDi08} 
for the film geometry.
% These theoretical predictions
% were compared with experimental results for the thermodynamic Casimir force
% between colloidal particles immersed in a binary liquid mixture and the 
% substrate.

As in ref. \cite{myCasimir}, where we had studied the strong adsorption limit,
we shall study the crossover by performing Monte Carlo simulations of
the improved Blume-Capel model on the simple cubic lattice. We shall give the
definition of this model in section \ref{definemodel} below.
Improved means that corrections to finite size scaling that are
$\propto L_0^{-\omega}$ vanish. This property is very useful in the study of
films, since typically the surfaces cause corrections $\propto L_0^{-1}$
\cite{BinderS,Diehl86,Diehl97} and fitting Monte Carlo data, it is quite 
difficult to disentangle corrections that have similar exponents.
Motivated by the experiments \cite{NeHeBe09}, we shall mainly study films 
where the external field $h_1$ at the first surface is finite, while
at the other surface the limit $h_2 \rightarrow \infty$ is taken, 
corresponding to the strong adsorption limit in a binary mixture.
For this choice of boundary conditions the correlation length of the
film divided by its thickness remains small at any temperature. In contrast,
for $h_1=h_2=0$ the film undergoes a second order phase transition in 
the universality class of the two-dimensional Ising model. This implies
that in the neighborhood of this transition the correlation length of the
film divided by its thickness is large. Therefore the Monte Carlo
study of the crossover from  $h_1=h_2=0$ to the limit 
$|h_1|, |h_2| \rightarrow \infty$ would be more involved than that performed
here.

In preparation for our study of the thermodynamic Casimir force, we 
have accurately determined the surface critical exponent $y_{h_1}$ of the 
ordinary surface universality class. Furthermore we have estimated the so called
extrapolation length for various boundary conditions. The extrapolation length
is directly related to the corrections to finite size scaling that are caused 
by the surfaces of the film.
Our numerical results are mainly based on the analysis of the behavior of the 
magnetization profile at the bulk critical temperature.
Next we have computed the thermodynamic Casimir force for the range 
of inverse temperatures around the bulk critical point where, at the level
of our numerical accuracy, it is non-vanishing. To this end we follow
the suggestion of Hucht \cite{Hucht}. For alternative methods see 
\cite{KrLa95,VaGaMaDi07,VaGaMaDi08,yet}.  Note that the stress tensor method
of \cite{DaKr04} can only be applied for periodic or anti-periodic boundary
conditions.  First we have simulated films
with a vanishing surface field $h_1=0$.  Based on the data obtained from these
simulations, we have also computed the Taylor-expansion of the 
thermodynamic Casimir force per area in $h_1$ up to the second order around 
$h_1=0$. We demonstrate that, taking into account corrections 
$\propto L_0^{-1}$, already for the relatively small thicknesses
$L_0=8.5$, $12.5$, and $16.5$ the behavior of the thermodynamic Casimir force 
per area as well as its partial derivatives with respect to $h_1$ is well 
described by universal FSS functions. Next we have simulated films with
various finite values of $h_1$ to check the range of applicability of the 
Taylor-expansion and to study the crossover beyond this range. 
Finally we have studied the approach to the strong adsorption limit 
$h_1 \rightarrow \infty$. Qualitatively we confirm the picture that emerges
from the exact solution of the two-dimensional Ising model \cite{AbMa09} and 
the mean-field calculation \cite{MoMaDi10}.

The outline of the paper is the following:  In section \ref{definemodel} we 
define the model and the observables that we have studied. In section 
\ref{semiinfite} we briefly review the phase diagram of a semi-infinite 
system. Then in section \ref{fsstheory} we discuss the finite size
scaling behavior of the magnetization profile at the bulk critical point 
and the finite size scaling behavior of the thermodynamic Casimir force.
In section \ref{computingCasimir} we discuss how to compute the thermodynamic
Casimir force and its partial derivatives with respect to the external field 
$h_1$  at the surface. In section \ref{MonteCarlo}
we present the results of our  Monte Carlo simulations. We performed 
a series of simulations at the bulk critical point, where we  focus on
the magnetization profile. Next we have determined the thermodynamic Casimir 
force per area in the neighborhood of the bulk critical point for various 
values of the external field $h_1$ at the surface.
Finally, in section \ref{Summary} we summarize and conclude.

\section{The model and bulk observables}
\label{definemodel}
We study the Blume-Capel model on the simple cubic lattice. 
It is characterized by the reduced Hamiltonian
\begin{equation}
 H = -\beta \sum_{<xy>}  s_x s_y + D \sum_x s_x^2  - h \sum_x s_x 
\end{equation}
where $x=(x_0,x_1,x_2)$ denotes a site of the lattice. The components 
$x_0$ ,$x_1$ and $x_2$ take integer values. The spin $s_x$ might take the  
values $-1$, $0$ or $1$. In the following we shall consider a vanishing 
external field $h=0$ throughout. The parameter $D$ controls the density of 
vacancies $s_x=0$. In the limit $D \rightarrow - \infty$ the spin-1/2 
Ising model is recovered. For $-\infty \le  D < D_{tri}$ the model undergoes 
a second order phase transition in the three-dimensional Ising
universality class. For $D> D_{tri}$ the transition is of first order.
The most recent estimate for the tri-critical point is
$D_{tri} = 2.0313(4)$ \cite{DeBl04}. 
Numerically, using Monte Carlo simulations it has been shown that there
is a point $(D^*,\beta_c(D^*))$ on the line of second order
phase transitions, where the amplitude of leading corrections to scaling 
vanishes.  Our most recent estimate is $D^*=0.656(20)$ \cite{mycritical}.
In \cite{mycritical} we have simulated the model at $D=0.655$ close to 
$\beta_c$ on lattices of a linear size up to $L=360$. From a standard finite
size scaling analysis of phenomenological couplings such as the Binder cumulant
we find 
\begin{equation}
\beta_c(0.655)=0.387721735(25) 
\end{equation}
for the inverse of the critical temperature at $D=0.655$.
The amplitude of leading corrections to scaling at $D=0.655$ is at least
by a factor of $30$ smaller than for the spin-1/2 Ising model. 

Our recent estimates for bulk critical exponents in the three-dimensional
Ising universality class are \cite{mycritical}
\begin{eqnarray}
 \nu &=& 0.63002(10) \;\;,\\
\eta &=& 0.03627(10) \;\;,\\
\omega &=& 0.832(6) \;\;.
\end{eqnarray}

In the following we set the scale by using the second moment correlation 
length $\xi_{2nd}$ in the high temperature phase of the model.  On a finite
lattice of the linear size $L$ in each of the directions it might be defined 
by
\begin{equation}
\label{xihigh}
\xi_{2nd}  = \sqrt{\frac{\chi/F-1}{4 \sin^2 \pi/L}} 
\end{equation}
where
\begin{equation}
F  =  \frac{1}{L^3} \left \langle
\Big|\sum_x \exp\left(i \frac{2 \pi x_k}{L} \right)
        s_x \Big|^2
\right \rangle
\end{equation}
is the Fourier transform of the correlation function at the lowest non-zero 
momentum and 
\begin{equation}
\chi  =  \frac{1}{L^3}
\left\langle \Big(\sum_x s_x \Big)^2 \right\rangle
\end{equation}
is the magnetic susceptibility. In \cite{myamplitude,myCasimir} we find 
\begin{equation}
\label{xi0}
\xi_{2nd,0,+} =  0.2282(2) - 1.8 \times (\nu-0.63002)
                        + 250 \times (\beta_c - 0.387721735)
\end{equation}
for the amplitude of the second moment correlation length in the high 
temperature phase, where we have used 
\begin{equation}
 t = \beta_c - \beta 
\end{equation}
as definition of the reduced temperature. We shall use this definition
of $t$ also in the following. The energy density is defined by
\begin{equation}
E_{bulk} = \frac{1}{L^3} \sum_{<xy>}  \langle  s_x s_y \rangle \;\;.
\end{equation}
In the following we shall need the energy density of the bulk system 
in a neighborhood of the bulk critical point. To this end, we 
have performed simulations at 350 different values of $\beta$ in the 
range $0.25 \le \beta \le 0.6$ \cite{myamplitude}. In a small neighborhood
of $\beta_c$, where no direct simulations are available we use
\begin{equation}
\label{critical2}
 E_{bulk}(\beta) = E_{ns} + C_{ns} (\beta-\beta_c)
             + a_{\pm} |\beta-\beta_c|^{1-\alpha}
             + d_{ns} (\beta-\beta_c)^2
             + b_{\pm} |\beta-\beta_c|^{2-\alpha} \;\;.
\end{equation}
For a discussion see section IV  of \cite{myamplitude}. 

\subsection{Film geometry and boundary conditions}
\label{definefilm}
Here we study systems with a film geometry.
In the ideal case this means that the system has a finite
thickness $L_0$, while in the other two directions the thermodynamic
limit $L_1, L_2 \rightarrow \infty$ is taken. In our  Monte Carlo
simulations we shall study lattices with $L_0 \ll L_1=L_2=L$ and
apply periodic boundary conditions in the $1$ and $2$ directions.
%Throughout we shall simulate lattices with $L_1=L_2=L$.

The reduced Hamiltonian of the Blume-Capel model with film geometry is  
\begin{eqnarray}
\label{Isingaction2}
H &=& - \beta \sum_{<xy>}  s_x s_y + D \sum_x s_x^2   \\
   &-& \beta_1 \sum_{<xy>,x_0=y_0=1}  s_x s_y
\;-\; \beta_2 \sum_{<xy>,x_0=y_0=L_0}  s_x s_y
\;-\; h_1 \sum_{x,x_0=1} s_x
\;-\; h_2 \sum_{x,x_0=L_0} s_x  \nonumber
\end{eqnarray}
where $h_1, h_2 \ne 0$ break the symmetry at the surfaces that are located
at $x_0=1$ and $x_0=L_0$, respectively.  In our convention $<xy>$ runs over 
all pairs of nearest neighbor sites with fluctuating spins. Note 
that here the sites $(1,x_1,x_2)$ and $(L_0,x_1,x_2)$ are not nearest 
neighbors as it would be the case for periodic boundary conditions.
In our study, 
we set $\beta_1=\beta_2=0$ throughout. Hence there is no enhancement of 
the coupling at the surface. There is ambiguity, where one
puts the boundaries and how the thickness of the film is precisely 
defined. Here we follow the convention that $L_0$ gives the number 
of layers with fluctuating spins.
In our previous work \cite{myCasimir} we have studied the limit of strong 
adsorption,  $|h_1|, |h_2| \rightarrow \infty$. In this  limit the spins
at the boundary are fixed to either $-1$ or $+1$. Therefore we had 
put the fixed spins on $x_0=0$ and $x_0=L_0+1$ to get $L_0$ layers
with fluctuating spins. Note that these fixed spins could also 
be interpreted as  external fields $h_{1,2} = \pm \beta$  acting on 
the spins at $x_0=1$ and $x_0=L_0$, respectively.
In the following we shall denote the type of 
boundary conditions by $(h_1,h_2)$. In the literature the cases $h_1=0$ or 
$h_2=0$ are often called free boundary conditions. To be consistent with the 
literature, we shall denote the strong adsorption limit by $+$ or $-$ 
in the following. In particular the two cases studied in \cite{myCasimir} are
denoted by $(+,+) \equiv (\beta,\beta)$ and $(+,-) \equiv (\beta,-\beta)$.
For the discussion
of the behavior of physical quantities near the boundary it is useful
to define the distance from the boundary. To this end we shall assume
that the first boundary is located at $x_0 = 1/2$ and the second one
at $x_0=L_0+1/2$. Hence the distance from the first boundary is
given by $z=x_0-1/2$ and the distance from the second one by
$z=-x_0 +L_0 +1/2$. 

In order to determine the thermodynamic Casimir force we have measured the 
energy per area of the film. It is given by
\begin{equation}
\label{energy}
 E = \frac{1}{L^2}  \left \langle \sum_{<xy>} s_x s_y \right \rangle \;\;.
\end{equation}
Since the film is invariant under translations in $1$ and $2$ directions but 
not in $0$ direction, the magnetization depends on $x_0$. Therefore we define
the magnetization of a slice by
\begin{equation}
m(x_0) = \frac{1}{L^2} \left \langle \sum_{x_1,x_2}  s_x \right \rangle \;.
\end{equation}

\section{Phase diagram of a semi-infinite system}
\label{semiinfite}
Here we briefly recall the phase diagram of a semi-infinite Ising system as it
is discussed e.g. in the reviews  \cite{BinderS,Diehl86,Diehl97}. 
For the Blume-Capel model, we expect that for $D \lessapprox 2$
the qualitative features of the phase diagram remain unchanged since 
$D_{tri} = 1.966(2)$ \cite{SiCaPl06} for the two-dimensional system
and $D_{tri} = 2.0313(4)$ \cite{DeBl04} for the three-dimensional one.

\begin{figure}[tp]
\includegraphics[width=13.3cm]{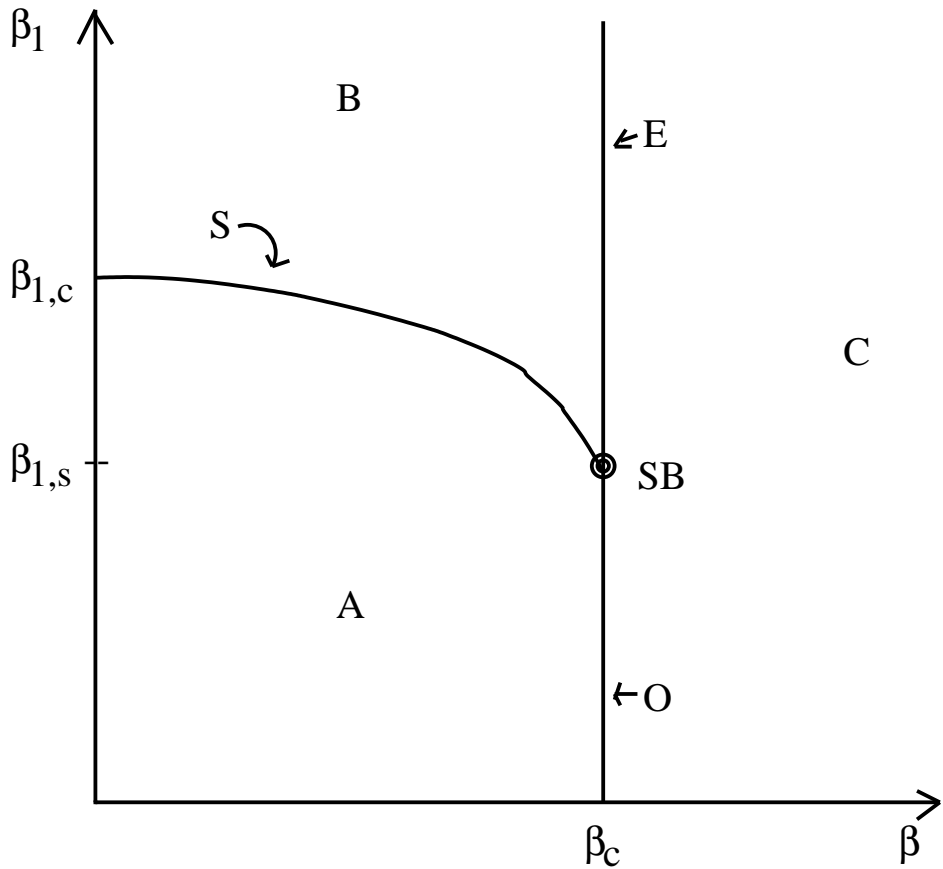}
\caption{\label{diagramm}
Sketch of the phase diagram of the semi-infinite system. 
On the $x$-axis we plot the 
coupling $\beta$ of the bulk and on the $y$-axis the excess coupling $\beta_1$
of the surface. A detailed discussion is given in the text.}
\end{figure}

In figure \ref{diagramm} we have sketched the phase diagram  
for a vanishing external field $h=0$ and a vanishing 
surface field $h_1=0$. For $\beta > \beta_c$ the spins in the bulk are ordered.
As a consequence, also the spins at the surface are ordered. This phase 
is denoted by C in figure  \ref{diagramm}. At vanishing bulk coupling
$\beta=0$ the spins at the surface decouple completely from those of the bulk.
Hence a two dimensional Ising or Blume-Capel model remains that undergoes 
a phase transition at $\beta_1=\beta_{c,2D}$. Starting from the point
$(0, \beta_{c,2D})$ there is a line of transitions, where the spins at the 
surface order, while those of the bulk remain disordered. This line 
hits the vertical line at $\beta=\beta_c$ in the so called special 
or surface-bulk point that we denote by SB in figure \ref{diagramm}, 
which is a tri-critical point.
In figure \ref{diagramm}, the phase, where both the boundary spins 
and those of the bulk are disordered is denoted by A while
the one with disordered bulk and ordered surface is denoted by B. The 
transitions from phase A to phase C are so called ordinary transitions,
while those from phase B to phase C  are so called extraordinary transitions.
The transitions from phase A to B are so called surface transitions.

For $h_1 \ne 0$ the spins at the surface are ordered also for 
$\beta_1 \le \beta_{1,s}$.  In the literature the transitions from disordered
to ordered spins in the bulk for $h_1 \ne 0$ % and $\beta_1 \le \beta_{1,s}$
are called normal transitions. In \cite{BuDi94}  it has been shown that
the normal surface universality class is equivalent to the extraordinary
surface universality class.

At the ordinary transition the external field $h_1$ at the surface 
is a relevant perturbation. Hence the RG-exponent $y_{h_1}$ associated with 
the surface field is positive.  In the literature, a number of surface
critical exponents have been introduced. In the case of the ordinary 
transition, these can be obtained from $y_{h_1}$ and the bulk RG-exponents
$y_t=1/\nu$ and $y_h = (d+2-\eta)/2$ by using scaling relations. 
In the following we need
\begin{eqnarray}
 \Delta_1 &=& \nu y_{h_1} \;\;,\\
 \beta_1  &=& \nu (d-1-y_{h_1})   \;\;,\label{betasc} \\
 \gamma_1 &=& \nu (2 -d/2 -\eta/2 + y_{h_1})  \;\;.
\end{eqnarray}
For the definitions and a complete list of these exponents see the reviews  
\cite{BinderS,Diehl86,Diehl97}. The numerical values of surface critical
exponents for the three-dimensional Ising universality class have been 
computed by various theoretical methods.
Mean field theory predicts $y_{h_1}=1/2$. The authors of \cite{BuEi78}
quote $y_{h_1} = 0.7363$ as result of their real space RG method and the 
authors of \cite{WhToGu79} quote $\gamma_1=0.78(2)$ as result of
a series expansion, which corresponds to $y_{h_1}=0.72(3)$.  
The $\epsilon$-expansion gives \cite{DiDi81}
\begin{equation}
y_{h_1}=\frac{1}{2} + \frac{1}{6} \epsilon + \frac{31}{321} \epsilon^2 
 + O(\epsilon^3) \;\;.
\end{equation}
Naively inserting $\epsilon =1$ one gets $y_{h_1}=0.666...$ and 
$y_{h_1}=0.762...$ at $O(\epsilon)$ and $O(\epsilon^2)$, respectively.
Using a massive field theory approach the authors of 
\cite{DiSh98} obtain $\Delta_1=0.45$ from the $[1/1]$ Pad\'e approximant of 
their two-loop result, which corresponds to $y_{h_1}=0.714$.  
Comparing  the different Pad\'e approximants that are given in table 9 
of \cite{DiSh98} one might conclude that the uncertainty of the estimate of
$y_{h_1}$ is about $0.02$. In table \ref{expotable} we have summarized
Monte Carlo results for surface critical exponents. Most of the 
authors quote an estimate for $\beta_1$ and some in addition for $\gamma_1$. In
those cases in which the authors did not quote a result for $y_{h_1}$ we have 
converted the value given for $\beta_1$ using the 
scaling relation~(\ref{betasc}) and $\nu=0.63002(10)$.

\begin{table}
\caption{\sl \label{expotable} 
Monte Carlo results for surface critical exponents for the 
ordinary phase transition in the three-dimensional Ising universality class.
The authors of \cite{KiOk85} quote no final result for $\gamma_1$. Here we 
give the average of the three results given in table II of \cite{KiOk85}. 
In case the authors do not quote an estimate for $y_{h_1}$, we have
computed it from $\beta_1$ and the scaling relation~(\ref{betasc}). 
These cases are marked by $^*$.
}
\begin{center}
\begin{tabular}{clll}
\hline
 Ref.           &  $\beta_1$    &  $\gamma_1$   &   $y_{h_1}$  \\   
\hline 
\cite{BiLa84}   & 0.78(2)   &           & 0.762(32)$^*$ \\
\cite{KiOk85}   & 0.79(2)   & 0.79(10)  & 0.746(32)$^*$ \\          
\cite{LaBi90}   & 0.78(2)   & 0.78(6)   & 0.762(32)$^*$ \\
\cite{NiBl93}   &           &           & 0.740(15)\\
\cite{RuWa95}   & 0.807(4)  & 0.760(4)  & 0.719(6)$^*$  \\
\cite{PlSe98}   & 0.80(1)   & 0.78(5)   & 0.730(16)$^*$   \\
\cite{DeBl03}   &           &           & 0.737(5)  \\
\cite{DeBlNi05} &0.796(1)   &           & 0.7374(15) \\
\cite{LiZh08}   &0.795(6)   &           & 0.738(10)$^*$  \\
\hline            
    here        &           &           &  0.7249(6)  \\
\hline            
\end{tabular}
\end{center}
\end{table}

For comparison we also anticipate our result for $y_{h_1}$ that we 
obtain in section \ref{p0bc}  below. Except for \cite{RuWa95} the 
estimates for $y_{h_1}$ are larger than ours. 
In particular, note that the difference 
between our result and that of \cite{DeBlNi05} is about six times as large
as the combined error.

\section{Finite size scaling  applied to films}
\label{fsstheory}
In this section we shall discuss the finite size scaling behavior of 
the magnetization profile at the bulk critical point and thermodynamic 
Casimir force for arbitrary temperature.
The starting point of our considerations is the reduced excess free energy 
per area of the film
\begin{equation}
\label{fex}
f_{ex}(L_0,t,h_1) = f_{film}(L_0,t,h_1) - L_0 f_{bulk}(t) 
\end{equation}
where $f_{film}(L_0,t,h_1)$ is the reduced free energy of the film 
per area and $f_{bulk}(t)$ the reduced bulk free energy density. There 
is no dependence on $h_2$, since we consider the limit $h_2 \rightarrow \infty$.
The singular part of the reduced excess free energy per area has the finite 
size scaling behavior \cite{BinderS,Diehl86,Diehl97}
\begin{equation}
\label{freemaster}
f_{ex,s}(L_0,t,h_1) = L_0^{-d+1} 
g(t [L_0/\xi_{0}]^{y_t}, h_1 [L_0/l_{ex,nor,0}]^{y_{h_1}})
\end{equation}
where we have ignored corrections to scaling at the moment and $d=3$ is the 
dimension of the bulk system. We shall define 
the amplitude $l_{ex,nor,0}$ of the normal extrapolation length $l_{ex,nor}$
below, eq.~(\ref{extralaw}). Note that the bulk contributions to the 
non-singular part of the free energy cancel in eq.~(\ref{fex}). However, 
there remain contributions from the two surfaces.

\subsection{The magnetization profile at the bulk critical point}
\label{fsstheoryM}
In terms of the reduced free energy per area the magnetization at $x_0=1$
is given by
\begin{eqnarray}
m_1 &=& \frac{\partial f_{ex}(L_0,t,h_1)}{\partial h_1}\nonumber
\\ 
&=& \frac{1}{L^2} \; \frac{1}{Z} 
\frac{\sum_{\{s\}} \exp(...+ h_1 \sum_{x_1,x_2} s_{(1,x_1,x_2)})}{\partial h_1}
=  \frac{1}{L^2} \left \langle \sum_{x_1,x_2}  s_{(1,x_1,x_2)} \right \rangle
\;\;.
\end{eqnarray}
In section \ref{p0bc} we shall determine the value of the RG-exponent 
$y_{h_1}$  from the scaling of $m_1$ with the thickness $L_0$ at $h_1=0$
and $\beta=\beta_c$. 
Taking the partial derivative of eq.~(\ref{freemaster}) with respect to $h_1$
we get
\begin{eqnarray}
m_1 &=& \left . \frac{\partial f_{ex}}{\partial h_1} \right |_{t=h_1=0} = 
 L_0^{-d+1} \left . 
\frac{\partial g(t [L_0/\xi_{0}]^{y_t}, h_1 [L_0/l_{ex,nor,0}]^{y_{h_1}})}
 {\partial h_1}  \right |_{t=h_1=0}
\nonumber \\
&=&L_0^{-d+1} \left . 
g_{h_1}(0,0) 
\right |_{t=h_1=0}
[L_0/l_{ex,nor,0}]^{y_{h_1}}  
= c \, L_0^{-d+1+y_{h_1}} 
\end{eqnarray}
where $g_{h_1}$ denotes the partial derivative of $g$ with respect to
$x_{h_1}=h_1 [L_0/l_{ex,nor,0}]^{y_{h_1}}$. 
Note that the non-singular contribution to $f_{ex}$ from the first surface
does not feel the breaking of the symmetry by the second surface. Therefore
it is an even function of $h_1$ and does not contribute to the partial
derivative with respect to $h_1$. 

The extrapolation length $l_{ex}$ can be defined by the behavior 
\cite{BinderS,Diehl86,Diehl97}
\begin{equation}
\label{magfinite}
 m(x_0) = c \, L_0^{-\beta/\nu} \; \psi(x_0/L_0)
\end{equation}
of the magnetization profile at the critical point of the bulk system.
Note that from scaling relations it follows that $\beta/\nu=(1+\eta)/2$, where
$\eta=0.03627(10)$ for the three-dimensional Ising universality class
\cite{mycritical}. 

In the neighborhood of the  surface with spins fixed to $s_x=1$, one 
expects that for $z \ll L_0$, where $z=L_0-x_0+1/2$, the magnetization 
profile does not depend on $L_0$. Therefore 
$\psi(x_0/L_0)=(z/L_0)^{-\beta/\nu}$ and hence \cite{BinderS,Diehl86,Diehl97}
\begin{equation}
\label{magfix}
 m(x_0) = c \, z^{-\beta/\nu}  \;.
\end{equation}
Also at the free boundary we expect that for $z \ll L_0$, where now
$z=x_0-1/2$, the functional
form of the magnetization profile does not depend on $L_0$. As we have seen
above, for a fixed value of $z$, the magnetization behaves as
$m_1 \propto L_0^{-d+1+y_{h_1}}$.  Therefore \cite{BinderS,Diehl86,Diehl97}
\begin{equation}
\label{magfree}
m(x_0) = a  \, z^{-\beta/\nu+d-1-y_{h_1}} = a  \, z^{(\beta_1-\beta)/\nu} \;\;.
\end{equation}
Since $-\beta/\nu < 0$, the scaling function of the magnetization profile 
diverges as $z/L_0 \rightarrow 0$ at the boundary with fixed spins. 
On the other hand since $(\beta_1-\beta)/\nu > 0$ the  scaling function of 
the magnetization vanishes as $z/L_0 \rightarrow 0$ at the free boundary.

Based on this observation one might define for finite thicknesses $L_0$ an 
effective distance from the boundary 
\begin{equation}
 z_{eff} = z + l_{ex} 
\end{equation}
such that the magnetization profile at $z_{eff}=0$ vanishes for $h_1=0$
or diverges in the case of symmetry breaking boundary conditions. The 
concept of the extrapolation length has been worked out explicitly for 
the ordinary transition in the framework of mean-field theory \cite{BinderS}.
Also in the Monte Carlo study of the magnetization profile of a semi-infinite
system in the extraordinary surface universality class an 
extrapolation length had been introduced \cite{SmDiLa94}. 
The extrapolation length is related with corrections $\propto L_0^{-1}$ 
discussed in the framework of field-theory in \cite{DiDiEi83}.
In the following we shall distinguish between the extrapolation length 
$l_{ex,ord}$ ($ord$ for ordinary surface transition) 
and $l_{ex,nor}$  ($nor$ for normal surface transition)
in the case of symmetry breaking boundary conditions. The extrapolation 
length depends on the precise definition of $z$. Physically, the 
extrapolation length depends on the details of the microscopic model,
in particular on the details of the fields and interactions at the surface.
In section \ref{extrapolation} we shall study the behavior of the 
extrapolation length as a function of the field $h_1$ at the boundary.
One expects  \cite{BrLei8283}
\begin{equation}
\label{extralaw}
 l_{ex,nor}(h_1) =  l_{ex,nor,0} \, h_1^{-1/y_{h_1}}
\end{equation}
which defines the amplitude $l_{ex,nor,0}$ that we have already used 
above in eq.~(\ref{freemaster}). 

Capehart and Fisher \cite{CaFi76} have argued that the arbitrariness in the 
definition of the thickness of the film leads to corrections $\propto L_0^{-1}$.
These corrections can be eliminated by replacing $L_0$ in finite size scaling 
laws such as eq.~(\ref{freemaster}) by an effective thickness
\begin{equation}
\label{leff}
L_{0,eff} = L_0 +L_s 
\end{equation}
of the film. Assuming that the corrections due to a surface are caused 
by a unique irrelevant surface scaling field, the constant $L_s$ should be 
given by 
\begin{equation}
\label{leffadd}
L_s =  l_{ex,1} + l_{ex,2} 
\end{equation}
where $l_{ex,1}$ and $l_{ex,2}$ are the extrapolation lengths at the 
two surfaces of the film. In section II A 4 of ref. \cite{Gaetal09} a similar 
discussion of the extrapolation length had been presented. 
For a discussion of the effective thickness and further references see
section IV of ref. \cite{myCasimir}.

\subsection{Crossover scaling function of the thermodynamic Casimir force}
In terms of the reduced excess free energy per area the thermodynamic Casimir 
force per area is given by \cite{Krech}
\begin{equation}
\label{defineFF}
\frac{1}{k_B T} F_{Casimir} = -  \frac{ \partial f_{ex}}{ \partial L_0} \;\;.
\end{equation}
Using the finite size scaling law~(\ref{freemaster}) we arrive at
\begin{eqnarray}
\frac{\partial f_{ex,s}(L_0,t,h_1)}{\partial L_0} &=& 
(-d+1) L_0^{-d} g(x_t,x_{h_1}) 
+ L_0^{-d} y_t t [L_0/\xi_0]^{y_t} \; g_t(x_t,x_h)  \nonumber \\
&+& L_0^{-d} y_{h1} h_1 [L_0/l_{ex,nor,0}]^{y_{h1}} g_{h_1} (x_t,x_{h_1}) 
\end{eqnarray}
where $x_t = t [L_0/\xi_0]^{y_t}$ and 
$x_{h_1} = h_1[L_0/l_{ex,nor,0}]^{y_{h1}}$.
The partial derivatives of $g$ with respect to $x_t$ and $x_{h_1}$ are 
denoted by $g_t$ and $g_{h_1}$, respectively. Note that the analytic part
of $f_{ex}$ is due to the surfaces and does not depend on $L_0$ and therefore
does not contribute to the thermodynamic Casimir force.
It follows that the thermodynamic Casimir force per area 
follows the finite size scaling law \cite{MoMaDi10}
\begin{equation}
\label{thetascaling}
 F_{Casimir} = k_B T L_0^{-d} 
   \Theta(t [L_0/\xi_0]^{y_t}, h_1 [L_0/l_{ex,nor,0}]^{y_{h1}})
\end{equation}
where 
\begin{equation}
 \Theta(x_t,x_{h_1}) = (d-1) g(x_t,x_{h_1})
  - y_t t [L_0/\xi_0]^{y_t} \; g_t(x_t,x_h)
  - y_{h1} h_1 [L_0/l_{ex,nor,0}]^{y_{h1}} g_{h_1}(x_t,x_{h_1}) \;\;.
\end{equation}
Taking the $n^{th}$ derivative of the thermodynamic Casimir force with 
respect to $h_1$ we get 
\begin{equation}
 \frac{\partial^n F_{Casimir}}{\partial h_1^n} = 
k_B T L_0^{-d} [L_0/l_{ex,nor,0}]^{n y_{h1}}
\frac{\partial^n \Theta(x_t, x_{h_1})}{\partial x_{h_1}^{n} } \;\;.
\end{equation}

\section{Computing the thermodynamic Casimir force and derivatives  
       with respect to the external field at the surface}
\label{computingCasimir}
On the lattice, we approximate the derivative of the reduced excess free energy
per area with respect to the thickness $L_0$ of the film by a finite difference:
\begin{equation}
\label{finitediff}
\frac{\partial f_{ex}}{\partial L_0} 
\approx \Delta f_{ex}(L_0) = f_{ex}(L_0+1/2) - f_{ex}(L_0-1/2) 
\end{equation}
where $L_0+1/2$ and $L_0-1/2$ are positive integers.
As suggested by Hucht \cite{Hucht} we compute this difference of free 
energies as the integral of the difference of corresponding internal 
energies:
\begin{equation}
\label{integralf}
\Delta f_{ex}(L_0,\beta) = - \int_{\beta_0}^{\beta} \mbox{d} \tilde \beta
                             \Delta E_{ex}(L_0, \tilde \beta)
\end{equation}
where 
\begin{equation}
\Delta E_{ex}(L_0) = E(L_0+1/2) - E(L_0-1/2) - E_{bulk} \;\;.
\end{equation}
In practice the integral~(\ref{integralf}) is computed by using the 
trapezoidal rule. Our previous experience \cite{myCasimir} shows that
$\Delta E_{ex}(L_0)$ has to be computed for about 100 values of $\beta$  
to obtain $\Delta f_{ex}(L_0,\beta)$ in the whole range of $\beta$ we are 
interested in at the level of accuracy we are aiming at.

In this work we compute the Taylor-expansion  of the thermodynamic Casimir
force with respect to the boundary field $h_1$ around $h_1=0$ up to the second 
order. To this end we compute the first and second derivative of
$\Delta f_{ex}(L_0)$ with respect to $h_1$. The $n^{th}$ derivatives can be
written as
\begin{equation}
\frac{\partial^{n} \Delta f_{ex}(L_0,\beta,h_1)}{\partial h_1^{n}} = 
-\int_{\beta_0}^{\beta} \mbox{d} \tilde \beta
 \frac{\partial^{n} \Delta E_{ex}(L_0, \tilde \beta,h_1)}{\partial h_1^{n}}
\end{equation}
where 
\begin{equation}
  \frac{\partial^n \Delta E_{ex}(L_0, \beta,h_1)}{\partial h_1^{n}}
= \frac{\partial^n E(L_0+1/2, \beta,h_1)}{\partial h_1^n}
- \frac{\partial^n E(L_0-1/2, \beta,h_1)}{\partial h_1^n} \;\;.
\end{equation}
Note that there is no bulk contribution, since the internal energy of the bulk 
does not depend on $h_1$. In the Monte Carlo simulation, the first derivative
can be computed as
\begin{equation}
\frac{ \partial E(L_0, \beta,h_1)}{\partial h_1} 
= \langle \tilde E M_1 \rangle - \langle \tilde E \rangle \langle M_1 \rangle 
\end{equation}
where 
\begin{equation}
 \tilde E = \frac{1}{L^2} \sum_{<xy>} s_x s_y 
\end{equation}
and 
\begin{equation}
 M_1 =  \sum_{x_1,x_2} s_{(1,x_1,x_2)}  \;\;.
\end{equation}
The second derivative is given by
\begin{equation}
\frac{\partial^2 E(L_0, \beta,h_1)}{\partial h_1^2}
 = \langle \tilde E M_1^2 \rangle 
 - 2 \langle \tilde E  M_1 \rangle \langle M_1 \rangle
  - \langle \tilde E  \rangle \langle M_1^2 \rangle
+ 2 \langle \tilde E \rangle \langle M_1 \rangle^2 \;\;. 
\end{equation}
Higher derivatives could be computed in a similar way. However it turns
out that the relative statistical error of the second derivative is much larger
than that of the first one. Therefore it seems  useless to implement and 
measure higher derivatives.

\section{Monte Carlo Simulation}
\label{MonteCarlo}
First we have simulated films with $(0,+)$ boundary conditions at the 
bulk critical point for thicknesses up to $L_0=64$. Analyzing the data
obtained from these simulations,
we have determined the value of $L_s$ for these boundary conditions and
have obtained an accurate result for the RG-exponent $y_{h_1}$. Next
we have simulated lattices of the size $L_0=L=512$ with $(+,0)$ and 
$(h_1,0)$ boundary conditions with $h_1=0.2$, $0.1$, $0.05$ and $0.02$ 
at the bulk critical point. From the behavior of the magnetization profile 
in the neighborhood of the surfaces we have 
determined the extrapolation length $l_{ex,ord}$ for free boundary conditions
and the extrapolation length $l_{ex,nor}$ as a function of $h_1$.
Next we have studied $(h_1,-)$ boundary conditions for $h_1=0.2$, $0.18$, 
$0.16$, $0.15$, $0.14$, $0.13$, $0.12$, $0.11$, $0.1$, $0.09$, $0.08$, $0.07$,
$0.06$ and $0.05$ also at the bulk critical point. From the zero of the 
magnetization profile, we read off the difference 
$l_{ex,nor}(h_1)-l_{ex,nor}(-)$ of extrapolation lengths. Note that  
$l_{ex,nor}(-)=l_{ex,nor}(+)$ due to symmetry.

Next we have studied the thermodynamic Casimir force per area 
in the neighborhood of 
the bulk critical point. To this end, we have simulated films of the 
thicknesses $L_0=8$, $9$, $12$, $13$, $16$ and $17$ for about $100$ 
values of $\beta$ each. Using the data obtained from these simulations
we have computed the finite size scaling function of the thermodynamic 
Casimir force per area for $(0,+)$ boundary conditions.
Furthermore we have computed the Taylor-expansion of the thermodynamic Casimir
force per area for $(h_1,+)$ boundary conditions to second order around 
$h_1=0$.  We have simulated $L_0=8$ and $9$ at $h_1=0.03$, $0.06$, $0.1$ and 
$0.2$ to check for how large values of $h_1$ and hence of $x_{h_1}$ the 
Taylor-expansion accurately describes the finite size scaling function
$\Theta(x_t,x_{h_1})$. Finally we have studied the approach to
the strong adsorption limit as $x_{h_1} \rightarrow \infty$. 

As in our previous work \cite{myCasimir} we have simulated the Blume-Capel
model by using a hybrid \cite{BrTa} of local heat-bath updates and 
cluster updates  
\cite{SW,Wolff}. Since the cluster updates only change the sign of the spins, 
additional local updates are needed to ensure ergodicity of the compound
algorithm. In one cycle of our algorithm we sweep twice through the 
lattice using the local heat bath algorithm followed by one or more 
cluster-updates. In one sweep  we run through the 
lattice in typewriter fashion, performing heat bath updates site by site.
We have always performed a cluster-update, in which all spins are flipped that 
are not frozen to the boundary. For a detailed discussion see section V A
of ref. \cite{myCasimir}. Note that here, in contrast to ref. \cite{myCasimir},
we have applied this type of cluster-update also to systems with $(+,-)$ 
boundary conditions. To this end we had to adapt the implementation 
of the cluster search; we had to allow for the possibility that two spins
in the cluster frozen to the boundary might have different signs.
Furthermore, we have generalized the cluster-update
to the case of a finite external field $h_1$ at the surface. A spin at the 
boundary freezes to the external field with the probability $p_f=1-p_d$, 
where 
\begin{equation}
p_d = \mbox{min}[1,\exp(-2 h_1 s_x)]  \;\;.
\end{equation}

In the case of large systems, discussed in sections 
\ref{sectionfree}, \ref{sectionnormal} below, we performed in addition
single cluster updates \cite{Wolff}. 
In all our simulations we have used the SIMD-oriented Fast Mersenne Twister 
algorithm \cite{twister} as pseudo-random number generator. 

\subsection{Simulations at the bulk critical point}
\label{p0bc}
First we have simulated films with $(0,+)$ boundary conditions 
at our estimate of the bulk critical point $\beta_c=0.387721735$ 
\cite{mycritical}. Since the fixed spins at the second surface act
effectively as an external field for the effectively two-dimensional 
system, the correlation length of the film stays finite at any value of 
$\beta$. This means that for a given thickness $L_0$, finite $L$ effects decay 
$\propto \exp(-L/\xi_{film})$ for sufficiently large values of $L$.
Hence we can chose $L$ such that finite size effects are much smaller than
the statistical errors and therefore can be ignored in the analysis 
of our Monte Carlo data.
In order to check which values of $L$ are needed to this end, we have
performed simulations for the thickness $L_0=6$, using
$L=6$, $7$, $8$, $9$, $10$, $11$, $12$, $13$, $14$, $15$, $16$,
$18$, $20$, $24$, $32$ and $48$. For each of these lattice sizes we have 
performed $10^9$ or more update cycles. As a check we have
simulated films with the thickness $L_0=12$ and 
$L=24$, $24$, $28$, $32$, $40$, $48$, $56$, $64$ and $96$, where
we performed $10^8$ update cycles throughout.
We have studied the behavior of the second moment correlation length,
the magnetization at the surface $m_1$,  and the energy per area of the film 
$E$ and its first and second derivative with respect to $h_1$.

\begin{figure}
\begin{center}
\includegraphics[width=13.5cm]{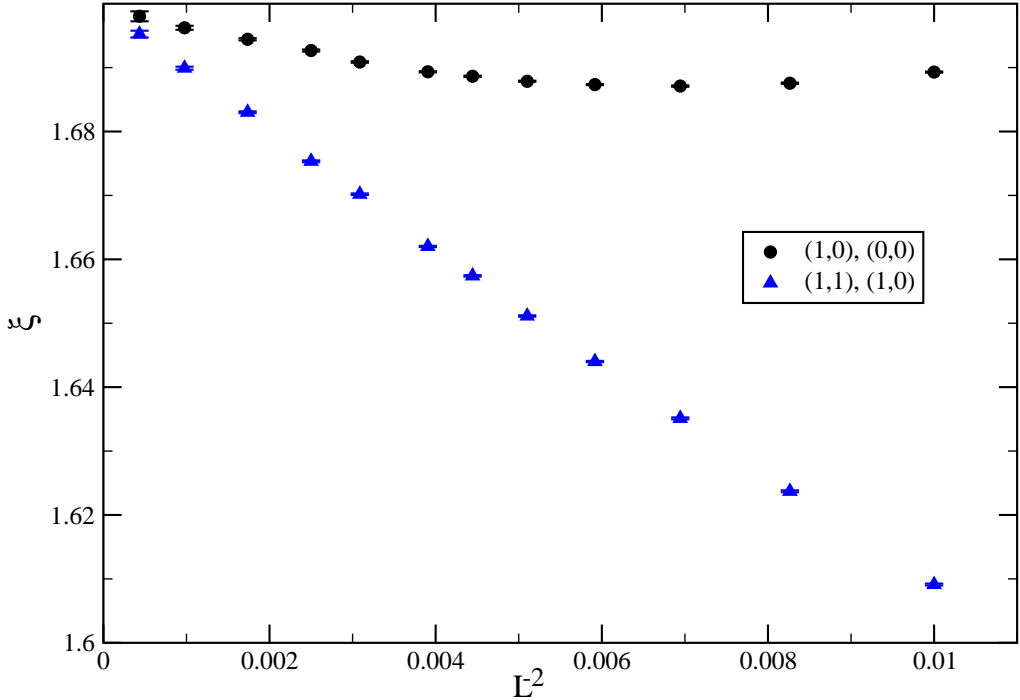}
\caption{\label{xi6}
We plot $\xi_{2nd}$  as a function of $L^{-2}$, where $L$ is the linear
extension in the transversal directions, for films of the  
thickness $L_0=6$. We have computed the second moment correlation length
by using the pairs of wave vectors $(\,(0,0), (1,0)\,)$ (circles) and 
$(\,(1,0), (1,1)\,)$ (triangles).
}
\end{center}
\end{figure}

Here we use the same definition of the second moment correlation length as
in ref. \cite{myCasimir}. See in particular section III C of \cite{myCasimir}.
The disadvantage of this definition of the second moment correlation length
is that as soon as more than one 
eigenstate of the transfer matrix contributes to the correlation function, 
corrections to the $L \rightarrow \infty$ limit only decay $\propto L^{-2}$. 
In figure \ref{xi6} we have plotted the second moment correlation 
length obtained with the pairs of wave vectors $(\,(0,0), (0,1)\,)$ 
and $(\, (1,0), (1,1)\,)$ as a function of $L^{-2}$.
While the estimate obtained by using the  pair of wave vectors 
$(\,(1,0), (1,1)\,)$ is monotonically increasing 
with increasing $L$, the estimate obtained by using the pair 
$(\,(0,0), (0,1)\,)$ displays
a minimum close to $L=12$. The value at this minimum is about $0.993$ times
the asymptotic value.
Fitting the results obtained for $L=24$, $32$ and $48$ with the ansatz 
$\xi_{2nd}(L) = \xi_{2nd} + a L^{-2}$ we get $\xi_{2nd} = 1.6988(6)$ 
and $1.6990(5)$ for the choices  $(\,(0,0), (1,0)\,)$  and 
$(\, (1,0), (1,1)\,)$, respectively.

Next we have analyzed the energy per area, its first and second derivative with
respect to $h_1$ and the magnetization $m_1$ 
at the surface. These quantities should converge with exponentially small
corrections as $L \rightarrow \infty$.  We have fitted these quantities 
with the ansatz $A(L) = A(\infty) + c_A \exp(-L/\xi_{film})$,
where we have taken our result for the second moment correlation length 
$\xi_{2nd}=1.70$, which should not be much smaller than the exponential 
correlation length that is actually needed here. Fitting all data with
$L \ge 16$ we find for the magnetization at the boundary
$\chi^2$/DOF $=4.55/4$, $m_1(\infty)=0.1250175(6)$ and $c_{m_1}=-0.237(25)$.
This means that for  $L \approx 11 \xi_{film}$ the deviation from the 
limit $L \rightarrow \infty$ has about the same size as the statistical 
error that we have reached here for $L_0=6$. Note that below, for larger 
thicknesses the number of measurements is more than a factor of ten smaller
than for $L_0=6$. Analyzing the energy per area and its first and second 
derivative with respect to $h_1$ we find that for $L \approx 10 \xi_{film}$
the deviation from the limit $L \rightarrow \infty$ has about the same size as 
the statistical error. Analyzing our results for the thickness $L_0=12$ we 
find consistently that for $m_1$ and the energy per area and its first and 
second derivative with respect to $h_1$, the deviation from the limit 
$L \rightarrow \infty$ has about the same size as the statistical error for 
$L \approx 10 \xi_{film}$. As we shall see below, 
$\xi_{film} \approx 0.225 (L_0+1.43)$. Therefore, for $L=4 L_0$, which
we have used below, the deviation from the limit $L \rightarrow \infty$ should
be clearly smaller than the statistical error and can hence be ignored.

Next we have simulated films for a large number of thicknesses up to 
$L_0=64$ at $\beta=0.387721735$, using $L=4 L_0$ throughout.
For the thicknesses 
$L_0=6$, $7$, $8$, $9$, $10$, $11$, $12$, $13$, $14$, $15$, $16$, $18$, 
$20$, $22$, $24$ we performed $10^8$ update cycles throughout, 
and $7.6 \times 10^7$, 
$10^8$, $8.7 \times 10^7$, $6.5 \times 10^7$, $4.3 \times 10^7$, 
$2.6 \times 10^7$, and $2.5 \times 10^7$ update cycles for $L_0= 28$, $32$,
$36$, $40$, $48$, $56$, and $64$, respectively.   
These simulations took about 18 months of CPU time on a single
core of a  Quad-Core AMD Opteron(tm) Processor 2378 running at 2.4 GHz.

We have fitted the data of the magnetization $m_1$ at the surface
with free boundary conditions with the ansatz
\begin{equation}
\label{m1fit1}
 m_1 = b \, (L_0 + L_s)^{2 - y_{h_1}}
\end{equation}
where $b$ and $y_{h_1}$ are the parameters of the fit and
\begin{equation}
\label{m1fit2}
 m_1 = b \, (L_0 + L_s)^{2 - y_{h_1}} \times [1 + c \, (L_0+L_s)^{-2}] 
\end{equation}
where now $c$ is an additional parameter,
to obtain some control on sub-leading corrections. We have  taken into 
account all data obtained for the thicknesses $L_0 \ge L_{0,min}$. 
Fitting our data with the 
ansatz~(\ref{m1fit1}) we get acceptable fits already for $L_{0,min}=10$:
$b=1.6131(13)$, $L_s=1.4289(33)$, $y_{h_1} = 0.72493(20)$, and 
$\chi^2/$DOF $=13.2/15$. Fitting with ansatz~(\ref{m1fit2}) we get
for $L_{0,min}=6$ the results 
$b=1.6109(22)$, $L_s=1.4166(86)$, $y_{h_1} = 0.72520(31)$, $c=-0.09(4)$, 
and $\chi^2/$DOF $=18.1/18$.

We arrive at the final estimates
\begin{eqnarray}
      b   &=& 1.613(4)
\label{b_est} \\
      L_s &=& 1.43(2)  
\label{Ls_est} \\
  y_{h_1} &=& 0.7249(6)
\label{yh_est}
\end{eqnarray}
where the central result is taken from the fit with the ansatz~(\ref{m1fit1}) 
and $L_{0,min}=10$. The error bar is chosen such that also 
the result for the fit with sub-leading corrections~(\ref{m1fit2}) is covered.
We have also estimated the error induced by the uncertainty 
of our estimate of the inverse bulk critical temperature $\beta_c$. 
To this end, we have first determined the derivative of $m_1$ with respect 
to $\beta$ for $L_0=8,9,12,13,16$ and $17$, where we performed simulations for 
many values of $\beta$.  We have extrapolated these results to other 
values of $L_0$ assuming 
$\partial m_1 /\partial \beta \propto (L_0+L_s)^{2-y_{h_1}+y_t}$. Using 
this we have computed $m_1$ at $\beta=\beta_c+\mbox{error}=0.38772176$ and 
have redone 
the fits performed above. We find that the deviations of the results for 
$\beta=0.38772176$ from those for $\beta=0.387721735$ are much smaller 
than the errors quoted in eqs.~(\ref{b_est},\ref{Ls_est},\ref{yh_est}).

Next we have analyzed the second moment correlation length obtained 
by using the pair $(\,(1,0), (1,1)\,)$ of wave vectors. Following the 
discussion above, finite $L$ effects might be still at the level of 
$1 \%$ for our choice $L=4 L_0$.  Since this effect is essentially the 
same for all thicknesses, it mainly effects the parameter $c$ in the 
two equations below.
First we have fitted our data with the ansatz
\begin{equation}
 \xi_{2nd} = c \, (L_0+L_s)  
\end{equation}
where $c$ and $L_s$ are the parameters of the fit. Using $L_{0,min}=8$ we
obtain $c=0.22435(9)$, $L_s=1.487(6)$ and $\chi^2/$DOF $=19.2/18$. 
Fitting instead with the ansatz
\begin{equation}
 \xi_{2nd} = c \, (L_0+L_s) \times [1 + b \, (L_0+L_s)^{-2} ] 
\end{equation}
we get for $L_{0,min}=6$ the results $c=0.22476(16)$, $L_s=1.422(20)$,
$b=0.48(12)$ and $\chi^2/$DOF $=15.0/19$. We conclude that the estimate 
for $L_s$ obtained from the finite size scaling behavior of $\xi_{2nd}$ 
is consistent with but less precise than that obtained from the 
finite size scaling behavior of $m_1$. 

Finally we have fitted the energy per area with the ansatz
\begin{equation}
 E = L_0 E_{ns} + E_{ns,s} + c \, (L_0+L_s)^{-2+1/\nu} 
\end{equation}
where we have used $E_{ns}=0.602111(1)$ \cite{myamplitude} and 
$\nu=0.63002(10)$ as input. 
Starting from our smallest thicknesses we get acceptable fits:
For $L_{0,min}=6$ we obtain $c=-3.5916(7)$, $L_s=1.4136(21)$, 
$E_{ns,s}=3.0644(2)$ and $\chi^2/$DOF $=18.8/19$.
As check we have also fitted with the ansatz
\begin{equation}
 E = L_0 E_{ns} + E_{ns,s} + c \, (L_0+L_s)^{-2+1/\nu} 
 \times [1+ b \, (L_0+L_s)^{-2}]
\end{equation}
where we have included sub-leading corrections.
For $L_{0,min}=6$ we get $c=-3.5930(15)$, $L_s=1.423(12)$, 
$E_{ns,s}=3.0646(2)$, $b=0.017(23)$ and $\chi^2/$DOF $=18.4/18$.

We have also redone the fits for shifted values of $E_{ns}$ and $\nu$. 
Since we have seen above in the case of $m_1$ that the 
uncertainty of $\beta_c$ is negligible, we have skipped this check here.
Taking all these results into account we arrive at the final estimates
\begin{eqnarray}
 E_{ns,s} & = & 3.064(1) \\
 L_s & = &  1.42(2)  \\
  c & = & 3.592(3) \;\;.
\end{eqnarray}
In particular we notice that the estimate of $L_s$ is fully 
consistent with that obtained above from the analysis of the 
magnetization $m_1$ at the boundary.  
In the following we shall use $L_s = 1.43(2)$ as obtained from the analysis
of the magnetization $m_1$ at the free boundary.

\subsection{The extrapolation length}
\label{extrapolation} 
First we have simulated lattices with $(h_1,0)$ boundary conditions 
of the size $L_0=L=512$  at $\beta = 0.387721735$  using 
$h_1=\beta_c$, $0.2$, $0.1$, $0.05$ and $0.02$. For this geometry one expects
strong finite $L$ effects.  However these should not alter the behavior in
the neighborhood of the boundary that we study here. In all cases we
have performed $2.6 \times 10^5$ update cycles.  
%For each measurement, we have
%performed two sweeps with the heat-bath algorithm, one cluster update in which 
%all spins that are not frozen to the boundary are flipped
%and $L_0$ single cluster updates. 
In total these simulations took about 7 months
of CPU time on a single core of a  Quad-Core AMD Opteron(tm) Processor 2378
running at 2.4 GHz.

\subsubsection{Behavior of the magnetization at the free boundary}
\label{sectionfree}
Following the discussion of section \ref{fsstheoryM} we have determined the 
extrapolation length $l_{ex,ord}$ by fitting our data for the magnetization
profile with the ansatz
\begin{equation}
\label{mOrdex}
 m(z) = c \, (z+l_{ex,ord})^{(\beta_1 - \beta)/\nu} 
\end{equation}
where $z$ gives the distance from the boundary as defined in section 
\ref{definefilm}, a few lines above eq.~(\ref{energy}).
To this end, we have computed the ratios
\begin{equation}
\label{ratioM}
 r(z) = m(z+1/2)/m(z-1/2) 
\end{equation}
to eliminate the constant in eq.~(\ref{mOrdex}). It  
turns out that cross-correlations of these ratios are relatively small. 
Therefore, for simplicity, we have fitted our data for these ratios 
taking only their statistical error into account, ignoring cross-correlations.
The statistical errors of the fit-parameters were computed by using 
a Jackknife procedure on top of the whole analysis, providing us with
correct statistical errors for the results.

\begin{figure}
\begin{center}
\includegraphics[width=13.5cm]{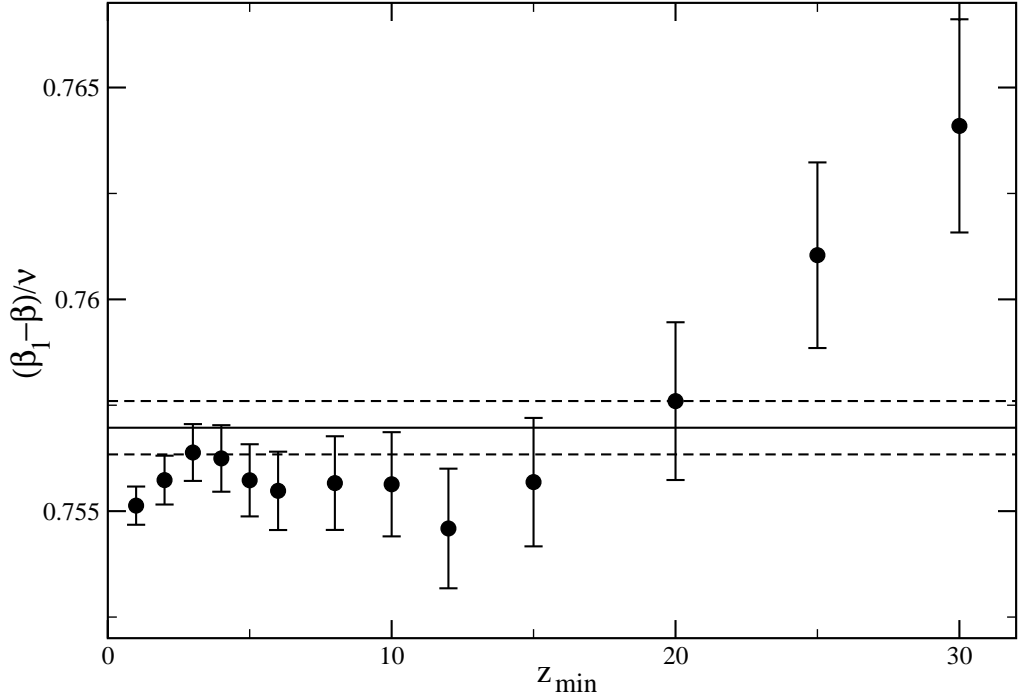}
\caption{\label{xvgl}
%%P9  x_eff eliminated
We plot the estimate of $(\beta_1-\beta)/\nu$ obtained by fitting with  
ansatz~(\ref{ansatzx}) as a function of $z_{min}$
(filled circles). In these fits distances $z_{min} \le z  \le 3 z_{min}$
from the boundary are taken into account. We have averaged the results over 
all values of $h_1$ that we have simulated. These results are compared with
$(\beta_1-\beta)/\nu=2-y_1-(1+\eta)/2=0.7570(7)$
obtained by using $y_{h_1}=0.7249(6)$, see the previous section, and 
$\eta=0.03627(10)$ \cite{mycritical}, where the central value is depicted
by the solid line and the error-bars are indicated by the dashed lines.
}
\end{center}
\end{figure}

First we have fitted our data with the ansatz 
\begin{equation}
\label{ansatzx}
r(z) = \left( \frac{z+l_{ex,ord}+1/2}{z+l_{ex,ord}-1/2} \right)^{(\beta_1-\beta)/\nu}
\end{equation}
where the free parameters of the fit are the extrapolation length $l_{ex,ord}$
and the exponent $(\beta_1 - \beta)/\nu$.
We have performed a large number of fits with various choices of the range
$z_{min} \le z \le z_{max}$ of distances from the boundary that are taken into
account, for all values of $h_1$ that we have simulated. The results for 
different $h_1$ are consistent among each other.
In figure \ref{xvgl} we show our results for the exponent $(\beta_1-\beta)/\nu$
for the choice $z_{max}=3 z_{min}$ as a function of $z_{min}$, where
we have averaged over all values of $h_1$ that we have simulated. The  error
that we give is purely statistical. For comparison we plot the  estimate of
$(\beta_1 - \beta)/\nu=2-y_{h_1} -(1+\eta)/2=0.7570(7)$ obtained by using our 
estimate of $y_{h_1}$, eq.~(\ref{yh_est}), and $\eta=0.03627(10)$ 
\cite{mycritical}.
We find that for $z_{min}=5$ up to $30$ the estimates obtained from the 
behavior of the magnetization profile in the neighborhood of the surface
are consistent with but less precise than the one using the estimate of 
$y_{h_1}$ obtained in the previous section. Therefore, in order to determine 
our final result for the extrapolation length $l_{ex,ord}$,  we have fixed
$(\beta_1-\beta)/\nu=0.7570(7)$. Fitting the data for $r(z)$ averaged over 
all values  of $h_1$ that we have simulated in the range  $5 \le z \le 30$ 
we arrive at
\begin{equation}
 l_{ex,ord} = 0.48(1) 
\end{equation}
where the error is dominated by the uncertainty of $(\beta_1-\beta)/\nu$.

\subsubsection{Normal extrapolation length as a function of $h_1$: part 1}
\label{sectionnormal}
Following the discussion of section \ref{fsstheoryM}  the magnetization in 
the neighborhood of the surface behaves as 
\begin{equation}
m(z,h_1) \propto (z + l_{ex,nor}(h_1))^{-\beta/\nu}  
\end{equation}
where $z$ gives the distance from the boundary. Also in the case of symmetry 
breaking boundary conditions, we have computed ratios~(\ref{ratioM})
of the magnetization of neighboring slices. These behave as
\begin{equation}
\label{rationor}
r(z) = \left( \frac{z+l_{ex,nor}+1/2}{z+l_{ex,nor}-1/2} \right)^{-\beta/\nu}
\;\;.
\end{equation}
Here we have solved eq.~(\ref{rationor}) with respect to $l_{ex,nor}$ for 
a single value of $z$, where we have used $\beta/\nu=0.518135$. For 
$h_1=\beta_c$ we find for $z \approx 15$ only a small dependence of the 
result on $z$. We read off $l_{ex,nor}=0.96(2)$.  In a similar way we have 
determined the extrapolation length for the other values of $h_1$. Our 
results are summarized in table \ref{Lextra0}. 

\begin{table}
\caption{\sl \label{Lextra0} The extrapolation length $l_{ex,nor}$ is obtained
for various values of $h_1$ by analyzing the behavior of the magnetization 
profile near the surface. For a discussion see the text.
}
\begin{center}
\begin{tabular}{ll}
\hline
 \multicolumn{1}{c}{$h_1$}
 & \multicolumn{1}{c}{$l_{ex,nor}$} \\
\hline
 $\beta_c$  &   0.96(2) \\
   0.2      &   2.25(3) \\
   0.1      &   5.56(4) \\
   0.05     &  14.0(2) \\
   0.02     &  $\approx 45$ \\
\hline
\end{tabular}
\end{center}
\end{table}

In ref. \cite{mycritical} we had determined $L_s=1.9(1)$ for $(+,+)$ 
and $(+,-)$ boundary conditions analyzing films of thicknesses up to $L_0=32$
at the critical point of the bulk system.
Now we have added for $(+,+)$ boundary conditions simulations for the 
thicknesses $L_0=48$, $64$ and $96$. This allows us to improve the accuracy
of our estimate. Now we find $L_s=1.90(5)$. This result is in perfect 
agreement with $L_s=2 l_{ex,nor}(\beta_c) = 1.92(4)$ obtained here.

For  $(0,+)$ boundary conditions we find $L_s=l_{ex,ord} + l_{ex,nor}(\beta_c)
= 0.48(1) + 0.96(2) = 1.44(3)$,  which is in perfect agreement with the result
given in eq.~(\ref{Ls_est}) above.

\subsubsection{Normal extrapolation length as a function of $h_1$: part 2}
%%P10 "$h_1 h_2 < 1$" to "$h_1 h_2 < 0$"
Here we have simulated systems with $h_1 h_2 < 0$, where $h_2=-\beta_c$, 
corresponding to fixed spins $s_x=-1$ at $x_0=L_0+1$, and various values 
of $h_1$.
For such a choice of boundary conditions the magnetization profile takes
positive values in the neighborhood of the first surface and negative 
ones in the neighborhood of the second surface.  Therefore, in between 
the magnetization profile vanishes at $x_{0,zero}$, which depends on 
$h_1$ and $h_2$.  The distance of this zero from the first boundary 
is given by $x_{0,zero}-1/2$ and from the second  boundary by 
$L_0-x_{0,zero}+1/2$.
Our basic assumption is that the zero of the magnetization indicates the 
physical middle of the system. Hence the distances of the zero from the 
effective positions of the first and the second boundary should be the same:
\begin{equation}
 x_{0,zero}-1/2 + l_{ex,nor}(h_1)  = L_0-x_{0,zero}+1/2 + l_{ex,nor}(h_2)
\end{equation}
and hence 
\begin{equation}
\Delta l_{ex,nor}(h_1,h_2)=l_{ex,nor}(h_1) -l_{ex,nor}(h_2) 
=L_0+1 -2 x_{0,zero} \;\;.
\label{getDelta}
\end{equation}
In order to define the zero of the magnetization we  have linearly interpolated
the magnetization profile. Throughout we simulate lattices with $L=4 L_0$.
First we have simulated at $h_1=0.2$, $0.1$ and $0.05$, using a large number 
of thicknesses $L_0$. Our results are summarized in table \ref{Lextra}. 
Apparently, $\Delta l_{ex,nor}$ converges with an increasing thickness $L_0$.
Numerically, corrections to the $L_0 \rightarrow \infty$ limit are 
compatible with an exponential decay. However we can not strictly exclude 
power-like corrections. For $h_1=0.2$ our results for $L_0 \ge 20$ are 
compatible within the statistical error. In the case of $h_1=0.1$ the results
for $L_0 = 40$ and $48$ are compatible. The one for $L_0=64$ is larger by 
about twice the combined statistical error than that for $L_0=48$. For 
$h_1=0.05$ the results for $L_0=120$ and $160$ are compatible. It is natural
to assume that the thickness $L_0$ needed to obtain $\Delta l_{ex,nor}$ 
with a given relative error is proportional to the extrapolation length
$l_{ex,nor}(h_1)$. Using $l_{ex,nor}(\beta_c) = 0.96(2)$ obtained in the 
section above, we conclude that for $L_0 \gtrapprox 10 l_{ex,nor}$ the 
deviation of $\Delta l_{ex,nor}$ from its $L_0 \rightarrow \infty$ limit
is less than the statistical error that we have reached here. Next we have
simulated at $h_1=0.18$, $0.16$, $0.15$, $0.14$, $0.13$, $0.12$, $0.11$, 
$0.09$, $0.08$, $0.07$ and $0.06$ for a single thickness $L_0$ each. 
The thicknesses $L_0$ and the estimates for $\Delta l_{ex,nor}$ are 
given in table \ref{Lextra}.  Throughout $L_0 > 10 l_{ex,nor}$ holds. 
For each of the simulations given in table \ref{Lextra} we performed about
$10^6$ update cycles. 
%Each update cycle consists of two sweeps with the 
%local heat-bath algorithm and one cluster-update, in which all spins are 
%flipped that are not frozen to the boundary. 
In total these simulations took about 8 months
of CPU time on a single core of a  Quad-Core AMD Opteron(tm) Processor 2378
running at 2.4 GHz.
Note that the results obtained here, are consistent with those of the 
previous section. Taking the numbers from table \ref{Lextra0}
we get $\Delta l_{ex,nor}=1.29(5)$, $4.60(6)$ and $13.04(22)$ for $h_1=0.2$,
$0.1$ and $0.05$, which is perfectly consistent with the results of the
present section, given in table \ref{Lextra}.

\begin{table}
\caption{\sl \label{Lextra} The difference of the extrapolation 
lengths $\Delta l_{ex,nor}(h_1,h_2)=l_{ex,nor}(h_1) -l_{ex,nor}(h_2)$,
where $h_2=-\beta_c$ as a function of $h_1$. For a discussion see the text.
}
\begin{center}
\begin{tabular}{lll}
\hline
 \multicolumn{1}{c}{$h_1$}
 & \multicolumn{1}{c}{$L_0$}
 & \multicolumn{1}{c}{$\Delta l_{ex,nor}$} \\
\hline
  0.2  &  10   & 1.2642(21) \\
  0.2  &  12   & 1.2748(25) \\
  0.2  &  14   & 1.2829(27) \\    
  0.2  &  16   & 1.2889(32) \\  
  0.2  &  18   & 1.2852(35) \\ 
  0.2  &  20   & 1.2955(38) \\  
  0.2  &  22   & 1.2989(42) \\  
  0.2  &  24   & 1.2938(44) \\ 
  0.2  &  28   & 1.2932(54) \\ 
  0.2  &  32   & 1.2941(62) \\ 
\hline
  0.18 &  28   & 1.6182(47) \\ 
\hline
  0.16 &  32   & 2.0434(61) \\ 
\hline
  0.15 &  36   & 2.313(7) \\  
\hline
  0.14 &  42   & 2.606(9)  \\ 
\hline
  0.13 &  48   & 2.969(9)  \\ 
\hline
  0.12 &  54   & 3.401(10)  \\ 
\hline
  0.11 &  60   & 3.925(11)  \\ 
\hline
  0.1  & 16  & 4.187(4) \\    
  0.1  & 20  & 4.330(5) \\  
  0.1  & 24  & 4.407(6) \\   
  0.1  & 28  & 4.461(6) \\  
  0.1  & 32  & 4.488(7) \\  
  0.1  & 40  & 4.546(8) \\  
  0.1  & 48  & 4.543(10) \\ 
  0.1  & 64  & 4.579(14) \\ 
\hline
  0.09 & 80  & 5.399(13) \\ 
\hline
  0.08 & 92  & 6.485(22) \\    
\hline
  0.07 & 110 & 7.917(25) \\ 
\hline
  0.06 & 140 & 9.892(35) \\   
\hline
  0.05 &  24 & 10.194(7) \\   
  0.05 &  32 & 11.108(9) \\    
  0.05 &  40 & 11.682(11) \\   
  0.05 &  48 & 12.060(13) \\   
  0.05 &  56 & 12.279(16) \\   
  0.05 &  64 & 12.492(17) \\   
  0.05 &  80 & 12.668(21) \\   
  0.05 & 120 & 12.892(29) \\  
  0.05 & 160 & 12.899(42) \\  
\end{tabular}
\end{center}
\end{table}

From eq.~(\ref{extralaw}) follows that
\begin{equation}
\label{Dl1}
\Delta l_{ex,nor} = l_0 + l_{ex,nor,0} |h_1|^{-1/y_{h_1}} 
\end{equation}
where naively $l_0=l_{ex,nor}(\beta_c)$.  However, since  
$l_{ex,nor}$ depends on the precise definition of the thickness 
of the lattice, we keep the offset $l_0$ as a free parameter here.

It turns out that for the range of $h_1$ that we have simulated here,  
analytic corrections have to be taken into account.  Therefore  
we have fitted our results for the difference of the extrapolation 
length with the ansatz
\begin{equation}
\label{Dl2}
\Delta l_{ex,nor} = l_0 + l_{ex,nor,0} |h_1 + a h_1^3|^{-1/y_{h_1}} 
\end{equation}
where the amplitude $l_{ex,nor,0}$ , the offset $l_0$ and the correction 
amplitude
$a$ are the parameters of the fit. Note that there should be no term 
$\propto h_1^2$ since $l_{ex,nor}(h_1)=l_{ex,nor}(-h_1)$. 
We set $y_{h_1}=0.7249(6)$ as obtained in section \ref{p0bc}.
In addition we have fitted with 
\begin{equation}
\label{Dl3}
 \Delta l_{ex,nor} = l_0 + l_{ex,nor,0} |h_1 + a h_1^3+b h_1^5|^{-1/y_{h_1}}
\end{equation}
to check for systematic errors due to the truncation of the 
Wegner expansion. Alternatively we have also fitted with  
\begin{equation}
\label{Dl2b}
 \Delta l_{ex,nor} = l_0 + l_{ex,nor,0} |h_1|^{-1/y_{h_1}} 
\times (1 + \tilde a h_1^2)
\end{equation}
and
\begin{equation}
\label{Dl3b}
 \Delta l_{ex,nor} = l_0 + l_{ex,nor,0} |h_1|^{-1/y_{h_1}} 
 \times (1 + \tilde a h_1^2 + \tilde b h_1^4) \;\;.
\end{equation}
Fitting with the ansaetze~(\ref{Dl2},\ref{Dl2b}) we get acceptable values of 
$\chi^2$/DOF starting from $h_{1,max}=0.2$, i.e. taking all data into account.
Discarding data with large $h_{1}$ the result for $l_{ex,nor,0}$ is slightly
decreasing and also $\chi^2$/DOF is further decreasing. E.g. fitting with  
ansatz~(\ref{Dl2}) and taking $h_{1,max}=0.14$ we get $l_{ex,nor,0}=0.2133(9)$,
$l_0=0.04(14)$, $a=6.9(2.2)$ and $\chi^2$/DOF $=1.72/7$.  Taking into account
the variation of the results over various ansaetze that we have used and 
the uncertainty of $y_{h_1}$ we arrive at 
\begin{equation}
 l_{ex,nor,0} = 0.213(3)
\end{equation}
which we shall use in the following.

\subsection{The thermodynamic Casimir force}
We have computed the thermodynamic Casimir force per area and its first and
second partial derivative with respect to $h_1$ for $(0,+)$ boundary 
conditions for the thicknesses $L_0=8.5$, 
$12.5$, and $16.5$. To this end we have simulated films of the thicknesses 
$L_0=8$, $9$, $12$, $13$, $16$ and $17$. For most of the simulations 
we have used $L=32$ for $L_0=8$ and $9$, $L=48$ for $L_0=12$ and $13$, and
$L=64$ for $L_0=16$ and $17$. The correlation length of the film displays
a single maximum at a temperature slightly below the critical temperature 
of the bulk system. The correlation length at the maximum is at most by 
one per mille  larger than at the critical point of the bulk system.
Therefore our choice of $L$ should ensure that finite $L$ effects of the
energy per area and its first and second partial derivative with respect to
$h_1$ can be safely ignored. At $\beta$-values that are much smaller or
larger than $\beta_c$ we have used smaller values of $L$. Throughout we have 
checked that $L > 10 \xi_{film}$ is fulfilled with a clear safety margin.
For $L_0=8$ and $9$ we have simulated at 85 values of 
the inverse temperature in the range $0.25 \le \beta \le 0.5$, for $L_0=12$ 
and $13$ at 124 values in the rage $0.3 \le \beta \le 0.42$, and for 
$L_0=16$ and $17$ at 112 values in the rage $0.34 \le \beta \le 0.406$.
The difference between neighboring $\beta$-values is adapted to the 
problem: It is the smallest close to $\beta_c$. We performed $10^8$ update 
cycles for $L_0=8$, $9$, $12$ and $13$ and $2 \times 10^8$ update cycles
for $L_0=16$ and $17$  for each value of $\beta$. 
In total these simulations took about 10 years of CPU time on a single core
of a Quad-Core AMD Opteron(tm) Processor 2378 running at 2.4 GHz.

\begin{figure}[tp]
\includegraphics[width=13.3cm]{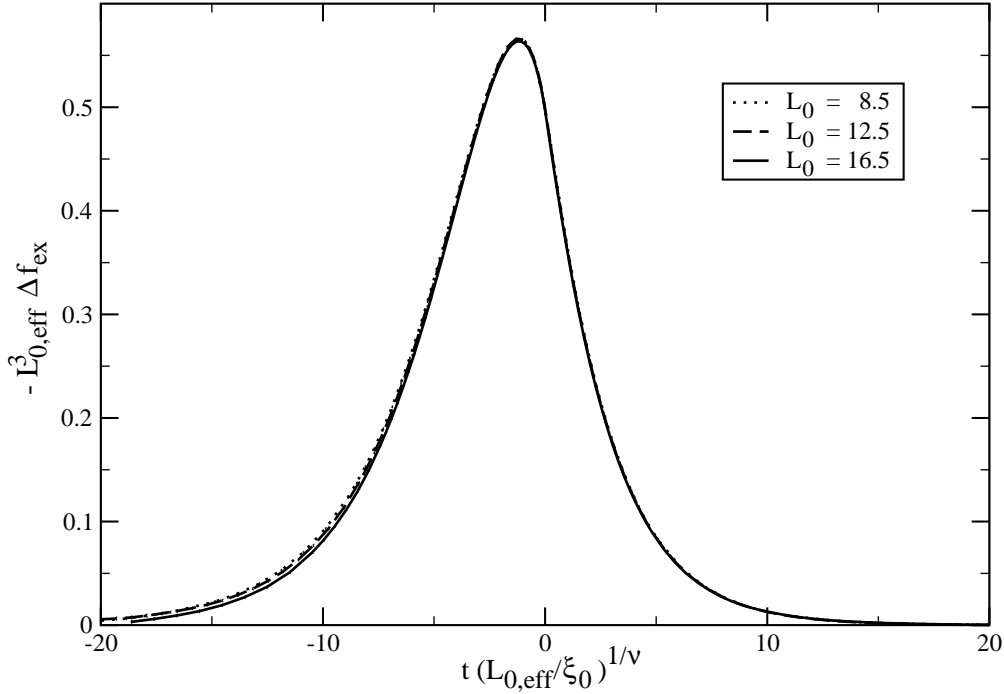}
\caption{\label{f+scaling} We plot $- L_{0,eff}^3 \Delta f_{ex}$ as
a function of $t (L_{0,eff}/\xi_0)^{1/\nu}$ for $(0,+)$ boundary conditions
for the thicknesses $L_0=8.5$, $12.5$ and $16.5$. To this end, 
we have used $L_{0,eff}=L_0+L_s$ with $L_s=1.43$, $\xi_0=0.2282$ and 
$\nu = 0.63002$. 
}
\end{figure}

Using the estimates of the energy per area obtained from these 
simulations we have computed the thermodynamic Casimir force per area 
as discussed in section \ref{computingCasimir}. In figure \ref{f+scaling}
we have plotted  $- L_{0,eff}^3 \Delta f_{ex}$ as a function of
$t (L_{0,eff}/\xi_0)^{1/\nu}$, where we have used $L_{0,eff}=L_0+L_s$ 
with $L_s=1.43$ obtained above in section \ref{p0bc}. We do not show
statistical errors in figure \ref{f+scaling}, since they are comparable with 
the thickness of the lines. The curves for
$L_0=8.5$, $12.5$ and $16.5$ fall quite nicely on top of each other. Only
for $x \lessapprox -7$, in the low temperature phase, we see a small 
discrepancy between the result for $L_0=8.5$, $12.5$ and $16.5$, which might
be attributed to analytic corrections. We conclude that we have obtained a good 
approximation of the finite size scaling function $\theta_{(0,+)}$. 

Throughout $\theta_{(0,+)}$ is 
positive, which means that the thermodynamic Casimir force is repulsive.
The scaling function $\theta_{(0,+)}$ has a single maximum. 
We have determined the position of this maximum from the zero of $\Delta E$.
We find $\beta_{max}=0.39069(2)$, $0.389443(10)$, and $0.388874(6)$ for
$L_0=8.5$, $12.5$, and $16.5$, respectively. It follows
$x_{t,max} = t_{max} (L_{0,eff}/\xi_0)^{1/\nu}=$ $-1.184(13)$, $-1.175(11)$, and
$-1.174(10)$ for $L_0=8.5$, $12.5$, and $16.5$, respectively.
The error bar includes the uncertainties of $\beta_{max}$, $L_s$ and 
$\nu$.  Note that the results obtained from the three different thicknesses
are consistent.
Next we have determined the value of the scaling function at the maximum.
We get $- L_{0,eff}^3 \Delta f_{ex}(x_{t,max})$ = $0.567(4)$, $0.566(3)$, and
$0.564(3)$ for $L_0=8.5$, $12.5$ and $16.5$, respectively.
The error is dominated by the uncertainty of $L_s$. The results
obtained from the three different thicknesses are consistent. As the final 
result we take the one obtained from $L_0=16.5$:
\begin{equation}
x_{t,max} = -1.174(10) \;\;, \;\;\; \theta_{(0,+),max} = 0.564(3) \;\;.
\end{equation}

At the critical point of the bulk system, the finite size scaling function
assumes the value
\begin{equation}
\theta_{(0,+)}(0) = 0.497(3) 
\end{equation}
where the error is dominated by the uncertainty of $L_s$.  This results 
can be compared with $\theta_{(0,+)}(0) = 0.33$, $0.416$ and $0.375(14)$ 
obtained by using the $\epsilon$-expansion, and Monte Carlo simulations
of the Ising model \cite{Krech97}. Similar to the case of $(+,+)$ and 
$(+,-)$ boundary conditions \cite{myCasimir}, we see a large deviations of 
the results of Krech from ours.

\begin{figure}[tp]
\includegraphics[width=13.3cm]{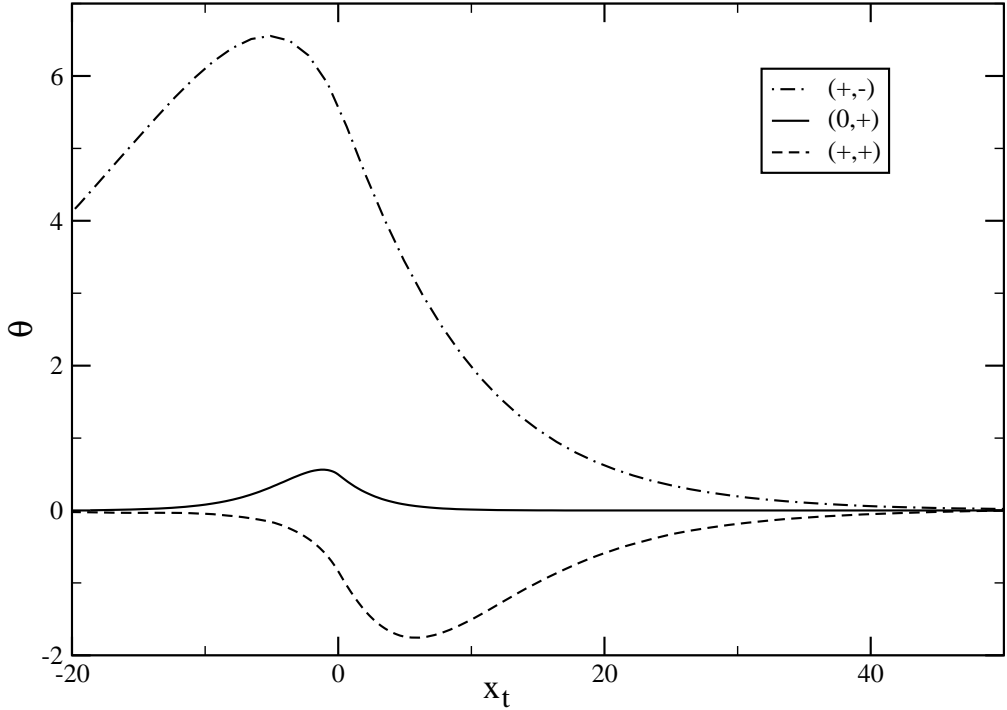}
\caption{\label{compare} We plot our result for the finite size scaling 
function $\theta_{(0,+)}$ along with those for $\theta_{(+,+)}$ and  
$\theta_{(+,-)}$ obtained in ref. \cite{myCasimir}.
}
\end{figure}

In figure \ref{compare} we compare the finite size scaling function of 
the thermodynamic Casimir force per area for $(0,+)$ boundary conditions 
with those of $(+,+)$ and $(+,-)$ boundary conditions that we have 
obtained in \cite{myCasimir}.

In the high temperature phase and around the bulk critical 
point, the absolute value of $\theta_{(0,+)}$  is smaller than that of
$\theta_{(+,+)}$, while in the low temperature phase for $x_t \lessapprox -1.1$
it becomes larger. The value of $\theta_{(0,+)}$ is much smaller than that of
$\theta_{(+,-)}$ throughout.

%%P12
As discussed in ref. \cite{Krech97}, see in particular eq.~(3.6) and the 
Appendix A of ref. \cite{Krech97}, in the mean-field approximation 
there is a simple relation between the scaling functions $\theta_{(+,-)}$ and 
$\theta_{(+,0)}$. For $(+,-)$ boundary conditions, the magnetization vanishes 
in the middle of the film. Hence, ignoring fluctuations, a film of the 
thickness $2 L_0$ with $(+,-)$ boundary conditions is composed of two 
films of the thickness $L_0$, where one has $(+,0)$ and the  other $(0,-)$
boundary conditions.  Furthermore $(0,+)$, $(+,0)$ and $(0,-)$ boundary 
conditions are equivalent. Therefore
\begin{equation}
\label{MFrelation}
 \theta_{MF,(0,+)}(x_t) = 2^{-d} \theta_{MF,(+,-)}(2^{1/\nu} x_t)  \;\;.
\end{equation}  
For less than four dimensions one expects deviations from this relation.  Indeed
for the Ising bulk universality class the ratio of Casimir amplitudes 
\begin{equation}
 \frac{\Delta_{(+,-)}} {\Delta_{(0,+)}} =
%          16 \frac{1-0.2822 \epsilon}{1+0.1988 \epsilon} =
          16 (1-0.481 \epsilon + ... )
\end{equation}
obtained by using the $\epsilon$-expansion \cite{Krech97}
clearly differs from $2^{4-\epsilon} = 16 (1 - 0.6931... \epsilon + ...)$
obtained from eq.~(\ref{MFrelation}). Note that the Casimir amplitude 
is given by $2 \Delta_{(b_1,b_2)} =  \theta_{(b_1,b_2)}(0)$. 
For two dimensions one obtains from conformal field theory \cite{Cardy86}
\begin{equation}
\frac{\Delta_{(+,-)}}{\Delta_{(+,0)}} = \frac{23}{2} 
\end{equation}  
which is almost 3 times as large as the factor $4$ predicted by 
eq.~(\ref{MFrelation}).

Taking our numerical data, we find  for $x_t>0$, this means in the 
high temperature phase,
$\theta_{(+,0)}(x_t) \approx 0.7 \times  2^{-3} 
                      \theta_{(+,-)}(2^{1/0.63002} x_t)$,
while in the low temperature phase, one gets
$\theta_{(+,0)}(x_t) \approx  2^{-3} \theta_{(+,-)}(2^{1/0.63002} x_t) - 0.3$
in the range $-10 < x_t < -3$. This means that eq.~(\ref{MFrelation}) does 
not provide a quantitatively accurate relation between the scaling functions
$\theta_{(+,0)}(x_t)$ and $\theta_{(+,-)}(x_t)$ in the three dimensional case.

%P13
%The discrepancy of the absolute value of $\theta_{(0,+)}$ on the one hand and
%$\theta_{(+,+)}$ and $\theta_{(+,-)}$ on the other is most striking in the 
%high temperature phase for $x_t \gg 0$. 
The most striking observation is that in the high temperature phase
$\theta_{(0,+)}$ decays, with increasing $x_t$, much faster to zero than
$\theta_{(+,+)}$ and  $\theta_{(+,-)}$ do. This behavior 
%of the thermodynamic Casimir
%force per area for $x_t \gg 0$ 
can be explained by using the transfer matrix formalism.
For a discussion of the transfer matrix formalism applied to the problem 
of the thermodynamic Casimir effect see section IV of \cite{myCasimir}. 
In terms of eigenvalues
$\lambda_{\alpha}$ and eigenvectors $| \alpha \rangle$ of the transfer matrix 
the thermodynamic Casimir force per area can be written as
\begin{equation}
\label{TCasimir}
\frac{1}{k_B T} F_{Casimir} =
- \frac{1}{L^2} \; \frac{\sum_{\alpha} m_{\alpha}  \exp(-m_{\alpha} l) \;
  \langle b_1 | \alpha \rangle \langle b_2 | \alpha \rangle}
  {\sum_{\alpha} \exp(-m_{\alpha} l)  \;
 \langle b_1 | \alpha \rangle \langle b_2 | \alpha \rangle} 
\end{equation}
where $1/\xi_{\alpha} = m_{\alpha} = - \ln(\lambda_{\alpha}/\lambda_0)$.
Note that here $m$ is a mass and should not be confused with the magnetization.
We assume that the eigenvalues are ordered such that 
$\lambda_{\alpha} \ge \lambda_{\beta}$ for $\alpha<\beta$, where $\alpha$, 
$\beta$ are positive integers or zero. The states $|b_1 \rangle$ and 
$|b_2 \rangle$ are defined by the boundary conditions that are applied and
$l=L_0+1$.  For $x_t \gg 0$ the right side of eq.~(\ref{TCasimir}) is dominated
by the contribution from the state $|1\rangle$ and therefore 
\begin{equation}
\tilde \theta_{(b_1,b_2)}(m l) 
\approx - m^3 l^3 \exp(-m l) C(b_1) C(b_2)
\end{equation}
where we have identified  $1/\xi=m = m_1$ and have defined
\begin{equation}
 C(b) = \frac{1}{m L} \frac{\langle b | 1 \rangle}
                           {\langle b | 0 \rangle} \;\;.
\end{equation}
The state $|0 \rangle$ is symmetric
under the global transformation $s_x \rightarrow -s_x$ for all $x$ in
a slice. Instead,  $|1 \rangle$ is
anti-symmetric and therefore $C=C(+) = - C(-)$.  It follows
\begin{equation}
\label{XZX}
\tilde \theta_{(+,+)}(m l) = - \tilde \theta_{(+,-)}(m l) =
- C^2 \; m^3 l^3 \exp(-m l)
\end{equation}
for sufficiently large values of $m l$. Since
$x_t = t [l/\xi_0]^{1/\nu} \simeq (m l)^{1/\nu}$ it follows
\begin{equation}
\label{scalinghigh}
\theta_{(+,+)}(x_t)=- \theta_{(+,-)}(x_t)= - C^2  x_t^{3 \nu} \exp(-x_t^{\nu})
\end{equation}
for sufficiently large values of $x_t$.
In the case of free boundary conditions the boundary state $|0\rangle$ is
symmetric under the global transformation $s_x \rightarrow -s_x$. Therefore
$\langle b |1\rangle$ vanishes and therefore 
\begin{equation}
  C(0) = 0  \;\;.
\end{equation}

\begin{figure}[tp]
\includegraphics[width=13.3cm]{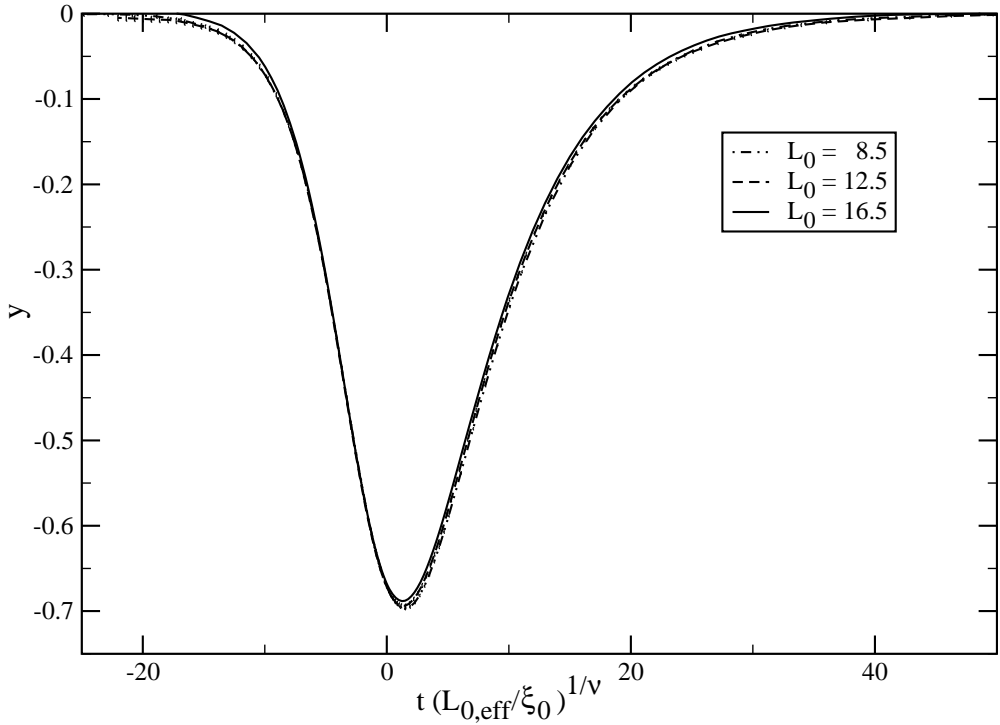}
\caption{\label{derive1} We plot  
$y=  - L_{0,eff}^3 (L_{0,eff}/l_{ex,nor,0})^{-y_{h_1}}
 \frac{\partial \Delta f_{ex}}{\partial h_1}$ as
a function of $ t (L_{0,eff}/\xi_0)^{1/\nu}$ for $(0,+)$ boundary conditions
for the thicknesses $L_0=8.5$, $12.5$ and $16.5$. To this end,
we have used $L_{0,eff}=L_0+L_s$ with $L_s=1.43$, $\xi_0=0.2282$ and
$\nu = 0.63002$.
}
\end{figure}

Next we have studied the first derivative of the scaling function 
with respect to $h_1$. In figure \ref{derive1} we have plotted 
$- L_{0,eff}^3 (L_{0,eff}/l_{ex,nor,0})^{-y_{h_1}} 
\frac{\partial \Delta f_{ex}}{\partial h_1}$ as a function of 
$t (L_{0,eff}/\xi_0)^{1/\nu}$. We do not give error bars, since the 
statistical error is of similar size as the thickness of the lines.
We find that the data for 
$L_0=8.5$, $12.5$ and $16.5$ fall quite nicely on top of each other. The 
small discrepancies that are visible for large absolute values
of $x_t$ might be attributed to analytic corrections. We conclude that
our numerical results provide a good approximation of the finite
size scaling function  
$\theta'(x_t) \equiv \left . \frac{\partial \Theta(x_t,x_{h_1})}{\partial h_1}
\right |_{h_1=0}$. 
We read off from figure \ref{derive1} that $\theta'$
is negative throughout and has a single minimum.

We have determined the location of this minimum
by searching for the zero of $\frac{\partial \Delta E}{\partial h_1}$. 
We find 
$\beta_{min} = 0.38403(3)$, $0.38577(2)$ and $0.38645(2)$ for 
$L_0=8.5$, $12.5$ and  $16.5$, respectively.  This corresponds to 
$x_{t,min} =  1.473(18)$, $1.333(18)$, and  $1.296(24)$. Here we have 
taken into account the errors of $\beta_{min}$, $L_s$ and $\nu$. In particular
for $L_0=16.5$ the error of $\beta_{min}$ clearly dominates. The results 
for $L_0=12.5$ and $16.5$ are consistent. % As final estimate we quote 
%$x_{t,min} = 1.30(5)$.  
As value of the derivative of the scaling function
we obtain $- 0.697(13)$, $-0.696(14)$ and $-0.688(13)$ for $L_0=$ $8.5$, 
$12.5$ and $16.5$, respectively. 
Note that in all cases about half of the error is due to the uncertainty in 
$l_{ex,nor,0}=0.213(3)$. 
The results for the different lattice sizes are consistent
within the quoted errors. We conclude
\begin{equation}
 x_{t,min} = 1.30(5) \;\;\;, \;\;\;   \theta'_{min} = -0.69(2) \;\;.
\end{equation}

Assuming that $C(h_1)$ is an analytic function and the finite size scaling 
behavior~(\ref{thetascaling}) of the thermodynamic Casimir force per area
we arrive at 
\begin{equation}
\label{scalinghighD}
\theta'(x_t) = B x_t^{3 \nu-\Delta_1} \exp(-x_t^{\nu})
\end{equation}
for $x_t \gg 0$.  Matching our numerical data for $L_0=16.5$ at 
$x_t \approx 10$ with eq.~(\ref{scalinghighD}) we arrive at $B=-0.85(5)$, 
where the error is estimated by comparing with the result obtained from 
$L_0=12.5$.

\begin{figure}[tp]
\includegraphics[width=13.3cm]{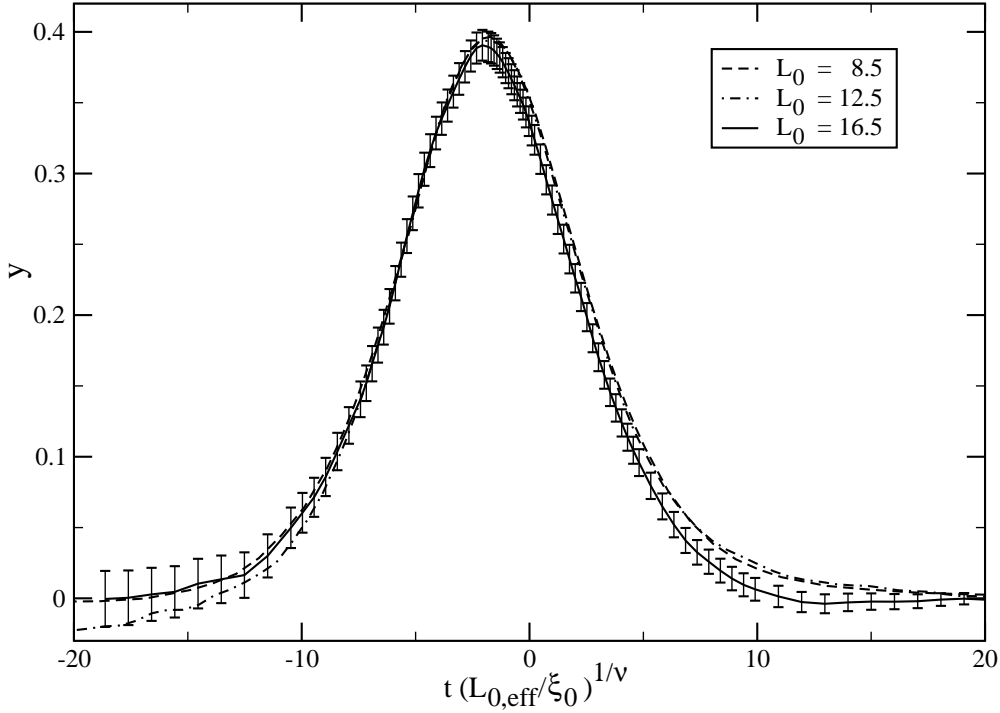}
\caption{\label{derive2} We plot 
$y=  - L_{0,eff}^3 (L_{0,eff}/l_{ex,nor,0})^{-2 y_{h_1}}
 \frac{\partial^2 \Delta f_{ex}}{\partial h_1^2}$ as
a function of $ t (L_{0,eff}/\xi_0)^{1/\nu}$ for $(0,+)$ boundary conditions
for the thicknesses $L_0=8.5$, $12.5$ and $16.5$. To this end,
we have used $L_{0,eff}=L_0+L_s$ with $L_s=1.43$, $\xi_0=0.2282$ and
$\nu=0.63002$.
}
\end{figure}

Next we have studied the second derivative of the scaling function
with respect to $h_1$. To this end, in figure \ref{derive2} we have plotted 
$- L_{0,eff}^3 (L_{0,eff}/l_{ex,nor,0})^{-2 y_{h_1}}
\frac{\partial^2 \Delta f_{ex}}{\partial h_1^2}$ as a function of
$ t (L_{0,eff}/\xi_0)^{1/\nu}$.  For $L_0=16.5$
we have plotted the statistical error, which we have not done for $L_0=8.5$, 
$12.5$ to keep the figure readable.
Within our statistical accuracy, the curves for the three
different thicknesses fall on top of each other. It seems that  $\theta''$ is 
positive for all values of the scaling function. Likely 
the negative values found for large $|x_t|$ and $L_0=16.5$ are just an 
artifact due to statistical fluctuations. 
The function displays a single maximum that is located at
\begin{equation}
 x_{t,min} = -1.9(2) \;\;\;, \;\;\;   \theta''_{min} = -0.39(2) \;\;.
\end{equation}

\begin{figure}[tp]
\vskip0.5cm
\includegraphics[width=13.3cm]{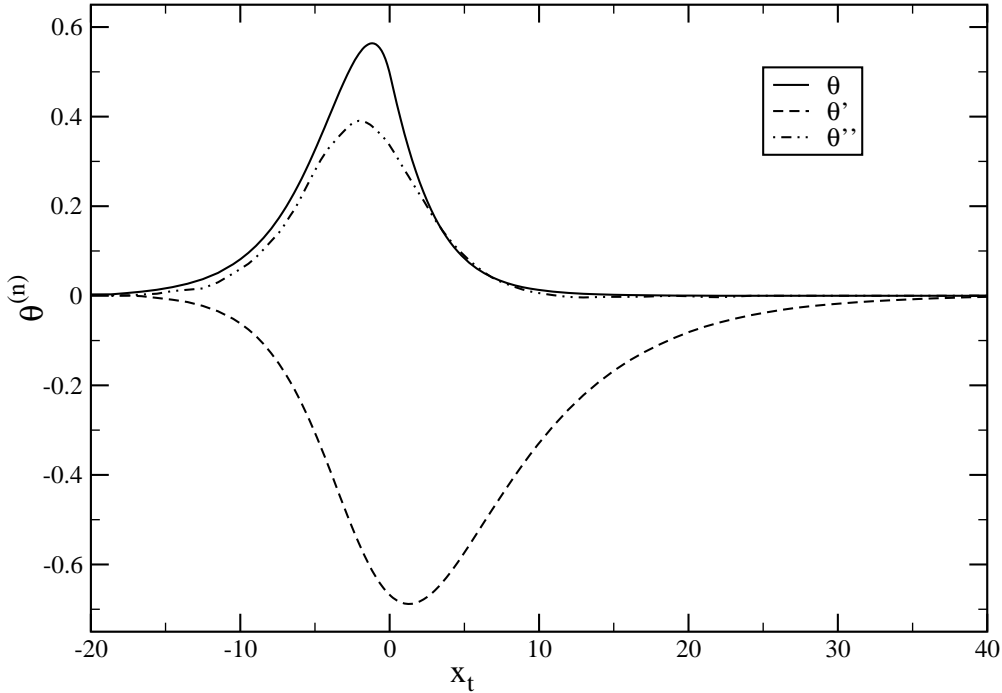}
\caption{\label{vglderiv}
We plot $\theta_{(0,+)}$, $\theta_{(0,+)}'$ and $\theta_{(0,+)}''$
as a function of $x_t$. 
}
\end{figure}

In figure \ref{vglderiv} we have plotted 
$\theta_{(0,+)}$, $\theta_{(0,+)}'$ and $\theta_{(0,+)}''$.  To this 
end we have used the results obtained for $L_0=16.5$.  
We find that the shape of $\theta_{(0,+)}''$ is quite similar to 
that of $\theta_{(0,+)}$. In particular, for $x_t \rightarrow \infty$,
both $\theta_{(0,+)}$ and $\theta_{(0,+)}''$  approach zero much faster
than $\theta_{(0,+)}'$. Therefore already for an infinitesimally small
positive value of $x_{h_1}$, the crossover scaling function 
$\Theta(x_t,x_{h_1})$, taken as a function of $x_t$, has a minimum in the 
high temperature phase.

\begin{figure}[tp]
\includegraphics[width=13.3cm]{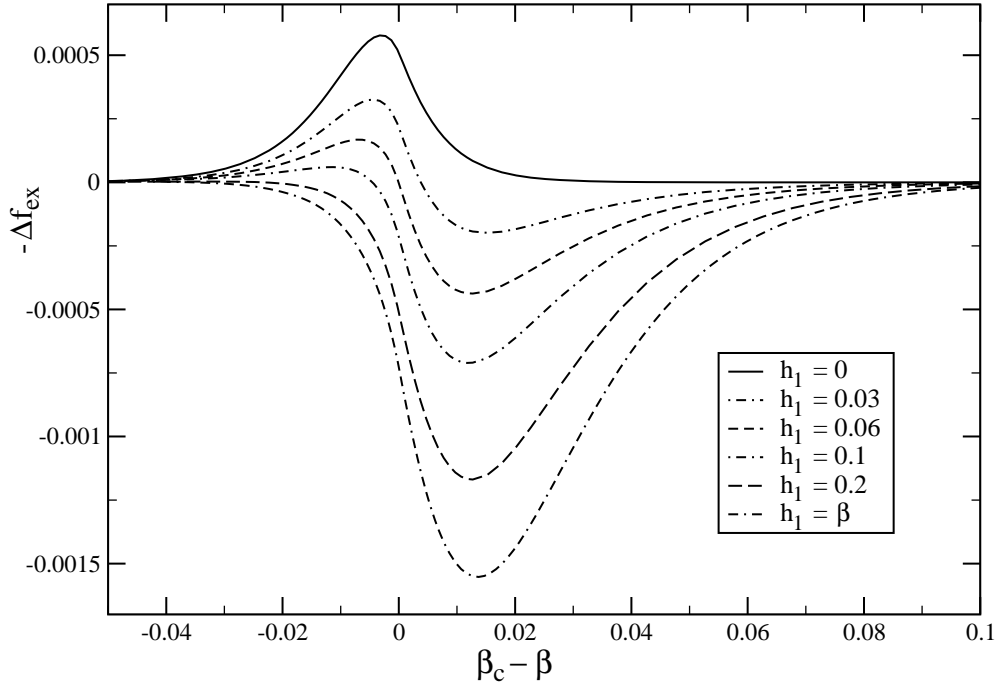}
\caption{\label{cross8}
We plot $- \Delta f_{ex}$ for $(h_1,+)$ boundary conditions
as a function of the reduced temperature $\beta_c-\beta$.
The thickness of the film is $L_0 = 8.5$ throughout.
}
\end{figure}

In order to check the range of applicability of the Taylor-expansion,
and to study the crossover beyond the Taylor-expansion, we have simulated 
films with $(h_1,+)$ boundary conditions and the thicknesses $L_0=8$ and $9$ 
at the values $h_1=0.03$,
$0.06$, $0.1$ and $0.2$ of the external field at the boundary. 
Our results along with that for $(+,+)$ 
corresponding to $h_1=\beta$ obtained in \cite{myCasimir} are plotted
in figure \ref{cross8}. For $h_1=0.03$ there is a minimum of the 
thermodynamic Casimir force per area in the high temperature phase. 
Its absolute 
value is about one third of the value of the maximum in the low temperature 
phase. The thermodynamic Casimir force changes sign at 
$\beta \approx 0.384$, which is slightly smaller than $\beta_c$.  Going 
to larger values of $h_1$ the position of the minimum changes only 
little and the absolute value of the minimum increases. On the other hand,  
the value of the maximum is decreasing with increasing $h_1$.  For $h_1=0.2$, 
the maximum has vanished.  

The authors of \cite{MoMaDi10} show in figure 9 of their paper 
Monte Carlo data obtained by O. Vasilyev \cite{Va} for the 
three-dimensional Ising model and the film thickness $L_0 = 10$.  There is 
nice qualitative agreement with our results given in figure \ref{cross8}.

We have compared the results for the thermodynamic Casimir force per area 
obtained by simulating at 
$h_1=0.03$, $0.06$, $0.1$ and $0.2$ for $L_0=8.5$ with those obtained by 
the Taylor-expansion around $h_1=0$ up to second order in $h_1$. We find 
that for $h_1=0.03$ the results almost agree within the statistical error.
Still for $h_1=0.06$ the Taylor-expansion to second order resembles the
true result quite well. The largest discrepancy is found for the value of 
the maximum of the thermodynamic Casimir force per area. 
It is overestimated by about a factor of 
$1.24$.  As one might expect, the result of the Taylor-expansion becomes 
increasingly worse with increasing $h_1$. In particular it does not reproduce
that for large values of $h_1$ the maximum of the thermodynamic Casimir force 
per area disappears.

Given our results for various thicknesses $L_0$ at $h_1=0$, we conclude that
the results for $L_0=8.5$ provide already a quite good approximation of the 
scaling limit.  In particular we are confident that the qualitative 
features of the crossover discussed here still hold in the scaling limit.
In particular we conclude that for 
$x_{h_1} \lessapprox 0.03 [(8.5+1.43)/0.213]^{0.7249} \approx 0.5$ 
the scaling function $\Theta(x_t,x_{h_1})$  is still well described
by the Taylor-expansion around $x_{h_1}=0$ to second order.

From figure \ref{cross8} we can read off that the thermodynamic Casimir
force can also change sign as a function of the thickness $L_0$ for 
fixed values of $h_1$ and the temperature. In general both 
$x_t = t [L_0/\xi_0]^{y_t}$ and $x_{h_1}= h_1 [L_0/l_{ex,nor,0}]^{y_{h_1}}$
depend on the thickness $L_0$. Therefore, for simplicity let us consider the 
bulk critical temperature, where $x_t=0$ for any thickness of the film. 
For small $L_0$ the scaling variable $x_{h_1}$ is small and therefore the 
thermodynamic Casimir force is close to the case $x_{h_1}=0$ and is therefore 
repulsive. As $L_0$ increases, $x_{h_1}$ increases and therefore 
$\Theta(0,x_{h_1})$ decreases. We read off from figure \ref{cross8} that
$\Theta(0,x_{h_1}) \approx 0$ for $x_{h_1} \approx 1$. With further increasing 
$L_0$, the thermodynamic Casimir force becomes attractive. 

\subsubsection{Approach to the $h_1 \rightarrow \infty$ limit}
For sufficiently large values of $x_{h_1} = h_1 (L_0/l_{ex,nor,0})^{y_{h_1}}$
we expect that corrections to the $x_{h_1} \rightarrow \infty$ limit can 
be described by replacing $L_0$ by $L_{0,eff}=L_0+L_s$, where 
\begin{equation}
\label{Lsadd}
L_{s} = l_{ex,nor}(h_1) + l_{ex,nor}(h_2) \;\;.
\end{equation}

\begin{figure}[tp]
\includegraphics[width=13.3cm]{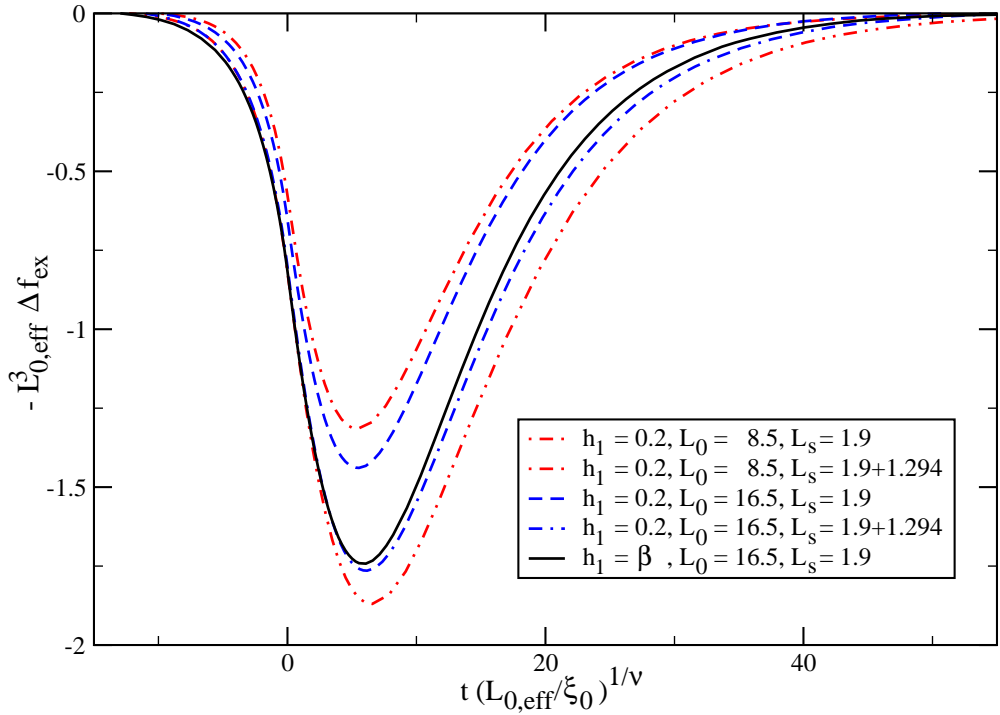}
\caption{\label{approach}
We plot $L_{0,eff}^3 \Delta f_{ex}$ as a function of 
$t (L_{0,eff}/\xi_0)^{1/\nu}$ for $h_1=0.2$ and $L_0=8.5$ and $16.5$ using
$L_s=1.9$  or $L_s=1.9+1.294$.
For comparison we give the corresponding curve for $L_0=16.5$ and $(+,+)$ 
boundary conditions using $L_s=1.9$. For a discussion see the text.
}
\end{figure}

In figure \ref{approach} we have plotted our results for $h_1=0.2$ and 
$L_0=8.5$ and $L_0=16.5$. First we use $L_s=1.9$ that we had obtained 
in ref. \cite{myCasimir} for $(+,+)$ boundary conditions and second 
$L_s=1.9+1.294$, 
where we have added $\Delta l_{ex,nor}(0.2,\beta_c)$ obtained in section
\ref{extrapolation}
above. For comparison we give the result obtained for $L_0=16.5$ and 
$(+,+)$ boundary conditions, using $L_s=1.9$. In the case of $L_0=8.5$ the
matching with the $(+,+)$ result is somewhat improved by using $L_s=1.9+1.294$
instead of $L_s=1.9$. While the value of the minimum is clearly improved,
the matching of the curve with that for  $(+,+)$ boundary conditions 
deep in the high temperature phase is not. 
In contrast, for $L_0=16.5$, using $L_s=1.9+1.294$ instead of $L_s=1.9$ 
clearly improves the matching of the curve for $h_1=0.2$ with that for
$h_1=\beta$ in the whole range of $x_t$  that is considered.

We conclude that for $L_0 \gtrapprox 10 \Delta l_{ex,nor}(h_1,\beta_c)$ using 
$L_s=1.9+ \Delta l_{ex,nor}(h_1,\beta_c)$ clearly improves the matching with 
the $(+,+)$ 
scaling function. It would be desirable to check this by simulations  for
smaller values of $h_1$. However this would be quite expensive, since already
for $h_1=0.15$ we would need to simulate a thickness $L_0 \approx 30$.  

\section{Summary and conclusions}
\label{Summary}
We have studied the crossover behaviors of a surface of a system in 
three-dimensional Ising universality class from the ordinary to the 
normal or extraordinary surface universality class. To this end, we 
have simulated the improved Blume-Capel model on the simple cubic lattice.
In particular we have studied films with various boundary conditions applied.
Improved means that corrections to finite size scaling $\propto L_0^{-\omega}$
have a vanishing amplitude,
where $L_0$ is the thickness of the film and $\omega=0.832(6)$ \cite{mycritical}
is the exponent of leading corrections. This property is very useful in the 
study of films, since corrections $\propto L_0^{-1}$ due to the surfaces 
are expected \cite{DiDiEi83} and fitting data it is difficult to 
disentangle corrections with similar exponents such as $\omega$ and one.
Mostly we have simulated films with $(0,+)$ boundary 
conditions. This means that at one surface we apply free 
boundary conditions, while at the other surface the spins are fixed to $+1$. 
Studying the magnetization of the slice at the surface 
with free boundary conditions, at the bulk critical point, of films
of a thickness up to $L_0=64$ we arrive at the estimate $y_{h_1}=0.7249(6)$
for the renormalization group exponent of the 
external field at the surface for the ordinary surface universality class.
This estimate is at least by a factor of 5 more accurate than those 
previously given in the literature. The authors of \cite{DeBlNi05} quote 
an error that is only 2.5 times larger than ours, however the deviation 
between our and their estimate is about 6 times larger than the combined 
errors. For details see table \ref{expotable}. We have studied the 
magnetization profile in the neighborhood of the surfaces for both the 
ordinary as well as the normal surface universality class. The data 
are consistent with the theoretically predicted power law behavior.
This study also allowed us to determine the extrapolation length $l_{ex}$ for 
free boundary conditions as well as symmetry breaking boundary conditions 
for various values of the external field $h_1$ at the surface. Corrections
to scaling $\propto L_0^{-1}$, which are due to the surfaces of the film 
can be expressed by an effective thickness $L_{0,eff} = L_0 + L_s$, where
$L_s$ depends on the details of the model.
Our numerical results confirm the hypothesis that $L_s=l_{ex,1} + l_{ex,2}$, 
where $l_{ex,1}$ and $l_{ex,2}$ are the extrapolation lengths at the two 
surfaces of the film.

Next we have studied the thermodynamic Casimir force in the neighborhood
of the bulk critical point in the range of temperatures where it does
not vanish at the level of our accuracy. First we have simulated films 
with $(0,+)$ boundary conditions and the thicknesses $L_0=8.5$, $12.5$
and $16.5$. Taking into account corrections by replacing  $L_0$ by $L_{0,eff}$,
the behavior of the thermodynamic Casimir force and its 
first and second derivative with respect to $h_1$ follows quite nicely 
the predictions of finite size scaling.  Hence our data allow us to 
compute good estimates of the finite size scaling functions $\theta_{(0,+)}$, 
$\theta_{(0,+)}'$, and $\theta_{(0,+)}''$. Next we
have computed the thermodynamic Casimir force per area for the thickness
$L_0=8.5$ at the finite values $h_1=0.03$, $0.06$, $0.1$  and $0.2$ of the 
external  field at the boundary. We find that the Taylor-expansion of the 
thermodynamic Casimir force up to the second order in $h_1$ around $h_1=0$
still describes the full function well at $h_1=0.03$ which corresponds to 
the value $x_{h_1} = h_1 [L_0/l_{0,ex,nor}]^{y_{h_1}} \approx 0.5$ of the 
scaling variable of the external field at the boundary. 
Finally, we have studied the approach of the thermodynamic Casimir force
to the limit $h_1 \rightarrow \infty$.  We find that by using 
$L_{0,eff}=L_0+L_s(h_1)$, the corrections to this limit are well described for
$L_0 \gtrapprox 10 [L_s(h_1)-L_s(\beta_c)]$. 

Based on  exact results for stripes of the two-dimensional Ising model
\cite{AbMa09}, mean-field calculations \cite{MoMaDi10} and preliminary 
Monte Carlo results for the Ising model \cite{Va} on the simple cubic lattice 
one expects that for certain combinations of the external fields $h_1$, $h_2$
and the thickness of the lattice $L_0$, the thermodynamic Casimir force 
changes sign as a function of the temperature. Also for 
certain choices of the external fields $h_1$, $h_2$ and the temperature,
the thermodynamic Casimir force changes sign as a function of the thickness 
of the film. Here we confirm these qualitative findings.

\section{Acknowledgements}
This work was supported by the DFG under the grant No HA 3150/2-1.

\end{document}